\documentclass{jfm}
\usepackage{graphicx}
\usepackage{savesym}
\usepackage{changes}
\savesymbol{comment}
\usepackage{epstopdf,epsfig,amsmath,comment}
\usepackage{caption}
\usepackage{subcaption}

\colorlet{Changes@Color}{blue}

\shorttitle{Fast-Tracking in Flight}
\shortauthor{S. A. Bollt, G. P. Bewley}

\author{Scott A. Bollt\aff{1,2},
  Gregory P. Bewley \aff{1}
  \corresp{\email{gpb1@cornell.edu}}}

\affiliation{\aff{1}Sibley School of Mechanical and Aerospace Engineering, Cornell University, 
Ithaca, NY 14850, United States
\aff{2}Graduate Aerospace Laboratories, California Institute of Technology, Pasadena, CA 91125, United States}

\title{How to Extract Energy from Turbulence in Flight by Fast Tracking}

\begin{document}

\maketitle




\begin{abstract}

We analyze a way to make flight vehicles harvest energy from homogeneous turbulence by fast tracking in the way that falling inertial particles do. 
Mean airspeed increases relative to flight through quiescent fluid when turbulent eddies sweep particles and vehicles along in a productive way. 
Once swept, inertia tends to carry a vehicle into tailwinds more often than headwinds. 
We introduce a forcing that rescales the effective inertia of rotorcraft in computer simulations. 
Given a certain thrust and effective inertia, we find that flight energy consumption can be calculated from measurements of mean particle settling velocities and acceleration variances alone, without need for other information. 
In calculations using a turbulence model, 
we optimize the balance between the work performed to generate the forcing and the advantages induced by fast tracking. 
The results show net energy reductions of up to about 10\% relative to flight through quiescent fluid 
and mean velocities up to 40\% higher. 
The forcing expands the range of conditions under which fast tracking operates by a factor of about ten. 
We discuss how the mechanism can operate for any vehicle, 
how it may be even more effective in real turbulence and for fixed-wing aircraft, 
and how modifications might yield yet greater performance. 




\end{abstract}

\section{Introduction}

A question central to the study of flight is the effect of flow unsteadiness 
on energy consumption. 
Range and endurance limit the utility of flight vehicles, 
particularly small ones 
\cite[][]{robobee,ChabotUAV,UAVreview}. 
While it is common to make predictions of range and endurance under the assumption that the air is quiescent, this assumption can be inaccurate. 
Given a specific trajectory, 
flight through unsteady air comes at the expense of the work performed to maintain the trajectory. 
Perhaps, the unsteadiness, or turbulence, can itself be so energetic that it represents an auxiliary energy reservoir that can be used to maintain flight. 
The challenge is to show if and when the latter case can prevail. 
Related questions apply to volant lifeforms \cite[][]{BirdBook,BirdEnergyPaper}. 

It is well-known that energy can be extracted from mean winds 
and large coherent structures in the atmosphere 
in order to extend range or endurance. 
Examples include thermal updrafts, mountain waves, and shear layers. 
These structures are approximately steady and predictable enough to be 
exploited by glider pilots \cite[][]{microlift,mountainwaves,LangelaanSUAVpath,CHUDEJhangglider}, 
birds \cite[][]{soaring2010,migrationsoaring}, 
and autonomous flight vehicles 
\cite[][]{MAVSoaringincities,soaringlikebirds,CFDcitySoaring,reddy2016learning}. 

Energy can also be extracted from the atmosphere when there is no mean wind 
by responding in specific ways to random gusts, or turbulent fluctuations. 
Birds such as the Albatross may do so 
\cite[][]{Pennycuick2002GustSA,Observedbirds,MallonBirdturbulence}. 
The majority of autonomous methods developed by humans to do so respond to flow measurements 
\cite[][]{Patel2006sine,Lissaman2007neutralcycles,MAVgustenergyextraction}, 
while birds or glider pilots may instead respond to their own accelerations  \cite[][]{MicroliftSoaringDescription,Kaseybirdturbulence}. 
\cite{birdgravityhorizon} shows that birds responded effectively to unsteady flows given even limited sensory information. 

\cite{Katzmayreffect} shows that fixed-wing aircraft can extract the energy in random gusts by clever transient rotations of the net aerodynamic force vector. 
To understand the effect, 
which \cite{Patel2009ExtractingEF} verifies in flight, 
consider that fixed-wing aircraft generally have much greater lift than drag so that their combination, or the net aerodynamic force on the aircraft, 
is almost normal to the direction of motion. 
Consequently, small upward gusts rotate the direction of the mean flow slightly in the reference frame of the wing and tilt the aerodynamic force forward transiently, which reduces drag (or increases thrust). 
Ignoring mean winds, upward and downward gusts are equally likely, but due to a nonlinearity the upward gusts cause larger net aerodynamic forces, so that transient drag reductions from upward gusts outweigh the corresponding increases from downward gusts. 
While gust velocities are smaller than the cruise speed of most aircraft, they are often on the same order as the downwash velocity so that vertical gusts can induce a significant change in the orientation of the lift vector relative to the aircraft's direction of flight, enough to cause flight power to drop transiently and even vanish \cite[][]{Pennycuick2002GustSA,Lissaman2007neutralcycles}. 
This makes vertical gust energy extraction effective for birds and fixed-wing aircraft. 
For rotorcraft, in contrast, the downwash velocity is typically large compared with vertical gust velocities so that flight power is not strongly affected. 
Neutrally buoyant vehicles do not require energy to maintain altitude (or depth for submarines) so that they cannot exploit the Katzmayr effect. 

The methods developed for fixed-wing aircraft as well as those employed by birds and glider pilots appear to have in common a tendency to amplify gust disturbances, in specific and controlled ways, rather than suppress them -- the opposite of what is typical in stability and control problems  \cite[][]{MicroliftSoaringDescription,Patel2009ExtractingEF,MallonBirdturbulence}. 
\cite{Glidergustamplification} notes that reducing glider inertia as well as adding positive feedback flaps to increase gust-induced accelerations can theoretically improve turbulent energy capture. 

Most algorithms for fixed-wing aircraft rely on the Katzmayr effect and the oversampling of flow in upwards gusts to extract energy from the gusts. 
Hence, these methods take a time-based signals approach to turbulence in the sense that the only necessary information about the flow is the vertical gust velocity as a function of time. 
The methods do not incorporate knowledge about the spatial structure of the flow. 
Rather than taking this approach, which results in appreciable benefits only for fixed-wing aircraft utilizing vertical gusts, gust energy capture has also been framed as a global path optimization problem. 
Given known wind fields, flight paths are routinely optimized to avoid headwinds and seek out tailwinds. 
With full knowledge of the flow, it is also possible to avoid downdrafts and seek out updrafts. 
These ideas apply underwater and on free surfaces as well, and are similar in spirit to updraft, thermal, and shear-layer soaring in that they typically only apply when flows are approximately stationary \cite[][]{LangelaanSUAVpath,underwaterglider,turbulenceavoidance,underwatercollisionandenergyopt,CHUDEJhangglider,underwaterpath}. 
The global approach to path optimization through turbulence is challenging because it requires rapidly updated flow-field measurements or real-time modeling and prediction of the flow. 
Furthermore, methods and algorithms employed at present on autonomous vehicles are often limited by the measurements the vehicle can itself make about its environment \cite[e.g.][]{UUVsensing}. 


In this paper we address the global path optimization problem using fluid dynamics to find efficient yet sub-optimal paths through turbulence without the need for real-time optimization algorithms. We analyze, theoretically, a way for vehicles to extract energy from turbulence by mimicking the aerodynamic coupling between inertial particles and turbulence. 
Inertial particles falling through turbulence 
naturally find nontrivial and energetically favorable paths that vehicles can follow using information only about their own accelerations, with no real-time modeling, and with only a parametric description of the flow. 
To see how this is possible requires an understanding of the way inertial particles behave in turbulence when gravity biases their direction of motion. 

Small particles fall down through turbulent flows faster on average than through a quiescent fluid; 
in some cases, nearly three times faster 
\cite[][]{Maxey87,Maxey87cell,WangandMaxey,Good,Bragg2019}. 
Though completely passive, 
particle find these favorable paths when their inertial timescale is resonant with a flow timescale, or in flows that evolve about as quickly as the particles can respond to this evolution. 
Under these conditions, particles tend to be swept toward the sides of vortices that push them down more quickly \cite[][]{WangandMaxey}. 

Rotorcraft, or any other vehicle, forced to act like particles of the right inertia can
passively find faster paths, 
albeit in the direction of their destination rather than toward the ground. 
To do so, a vehicle needs to apply forces proportional to its measured instantaneous accelerations, for instance, and thereby modify its effective mass so that it reacts to gusts just as fast-tracking particles do, but with a bias toward a destination provided by a mean thrust rather than by gravity. 
This results in energy extraction from turbulence in spite of a lack of knowledge about the instantaneous structure of the surrounding flow. 
It is a proof of this principle that we explore in this paper. 

We call the forcing cyber-physical since it 
changes the effective inertia of the rotorcraft. 
The concept of using cyber-physical tools to achieve desired interactions between a body and flow has been explored before. 
\cite{Williamsoncyber} for instance studies fluid-structure interactions and vortex shedding on a cylinder. 
Previous implementations rely on tethered force measurements rather than untethered acceleration measurements in their computations
\cite[][]{hover_techet_triantafyllou_1998,Williamsoncyber}. 

We focus in this paper primarily on rotorcraft that are smaller than the size of the dominant turbulent flow structures through which they fly, and that move in only two dimensions, 
one of which is in the direction of a destination. 
The two dimensions are perpendicular to gravity, 
with a mean thrust for rotorcraft playing the role of gravity for inertial particles. 

In Sec.~\ref{theorysection}, we review a simple model of rotorcraft flight
and propose a simple cyber-physical forcing on the rotorcraft. 
We find that the form of the dimensionless equations of motion is the same as the one for settling particles. 
The forcing allows 
rotorcraft to mimic a particle of any settling parameter and Stokes number. 
While the forcing allows any place in parameter space to be reached in principle, 
there is a cost to doing so determined by the magnitude of the forces the rotorcraft needs to generate in order to mimic the desired particle dynamics. 
We find that these costs are determined in part by the moments of the probability density function of inertial particle accelerations. 
The balance between the costs and the gains realized by moving into energetically favorable parts of parameter space lead to the existence of optimal shifts in the parameters, which depend on the characteristics of the turbulence and of the rotorcraft in ways that we calculate. 
The methods section (Sec.~\ref{methodssection}) describes how we simulated turbulence, 
rotorcraft flight, and how we perform the optimizations. 


In Sec.~\ref{resultssection} we present the advantages realized by a simple cyber-physical forcing, abbreviated FT. 
The purpose of the calculations is to delineate the boundaries in parameter space within which potential gains can be realized by the forcing. 
We find that compared with flight through quiescent flow (QF), fast-tracking forcing (FT) reduces both energy consumption and flight time. 
The advantages are significant for rotorcraft with natural response times faster than the characteristic turnover time of the flow, and for vehicles with cruising speeds within an order of magnitude of the characteristic speed of turbulent fluctuations in the flow. 
Relative to doing nothing (DN), in a sense explained below, the advantage of the forcing is to broaden the range of conditions under which turbulence benefits flight, particularly if the effective vehicle inertia is anisotropic as explained in the theory section. 
Doing nothing in turbulence is automatically beneficial relative to flight through quiescent flow due to intrinsic fast-tracking, provided the relevant dynamics apply or can be made to apply to a vehicle. 
The cost of gust suppression, or disturbance rejection (DR), is large compared with the gains realized by any other flight mode. 

We expect that further benefits to flight may be realized through increased sophistication of the forcing model, ideas for which we review in Sec.~\ref{discussionsection}. 
Furthermore, comparisons with experiments on particles in turbulence suggest gains up to ten times larger than those we found in our calculations \cite[][]{WangandMaxey,Good}. 
The reduced gains appearing in the calculations are comparable to those achieved in previous studies using turbulence models that respect turbulence statistics and kinematics but ignore the dynamics of real turbulence. 
This may be the result of vorticity in the models not being as strongly correlated spatially or temporally as in real turbulence. 
Finally, we believe that the theory can be generalized to three dimensions and to any properly forced vehicle moving in a turbulent fluid. 

\section{Theory}
\label{theorysection}
As a foundation for autonomous flight strategies to navigate turbulent flows efficiently, 
we use a simple model of flight vehicle dynamics to show how it leads naturally to a forcing strategy. 
The flight vehicle is a rotorcraft, meaning that the thrust not only propels the vehicle but also directly supports its weight. 
One component of the thrust points in a fixed direction, meaning that the destination for the flight vehicle is at $\infty$, or far away. 
We consider statistically homogeneous, isotropic turbulence with a zero mean, and for further simplicity, we consider flows that fluctuate only in the plane perpendicular to gravity. 
Fast tracking operates in both two and three dimensions, and we expect the results we observe in two dimensions to generalize to three \cite[][]{MaxeyCorrsin86,RosaDNS}. 
The potential advantages are realized statistically, meaning that our results are expectation values for many flights, or for long flights, through statistically stationary turbulent flows. 

\begin{figure}
  \centerline{\includegraphics[width=0.25\linewidth]{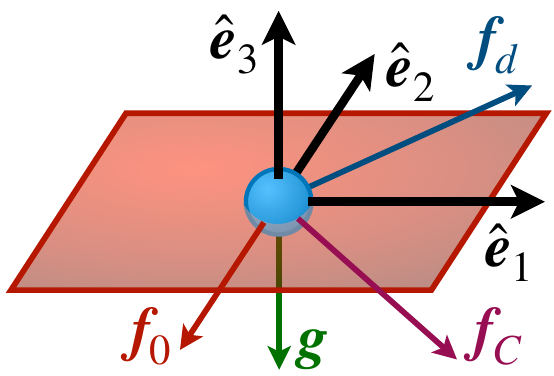}}
  \caption{Movement is in the $\hat{\boldsymbol{e}}_1$-$\hat{\boldsymbol{e}}_2$ plane (red), 
  while gravity (${\boldsymbol{g}}$) points in the $-\hat{\boldsymbol{e}}_3$ direction. 
  The constant component of the thrust, ${\boldsymbol{f}}_0$, points opposite to $\hat{\boldsymbol{e}}_2$, and additional components defined in the text 
  include the one given by the forcing, ${\boldsymbol{f}}_C$. 
  Drag on the vehicle, ${\boldsymbol{f}}_d$, depends on the relative velocity between the vehicle and fluid. 
}
\label{coordinates}
\end{figure}

We compare the case of flight through turbulence under fast-tracking forcing (FT) 
to the cases of flight through each of quiescent fluid (QF), 
turbulent fluid while doing nothing (DN), and turbulent fluid while rejecting disturbances (DR). 
The letters in parentheses appear as subscripts to denote the conditions under which different quantities were calculated. 
While the DN case does not correct deviations from its path caused by turbulence, implicit in all cases is the assumption that the rotorcraft controls its angular degrees of freedom quickly compared to the dynamics of interest; this may be a better assumption for rotorcraft than for fixed-wing aircraft in turbulence \cite[][]{rotorcraftvsfixedwinginturbulence}. 

\subsection{Particle Dynamics and Fast Tracking}
The momentum equation for heavy particles balances the particle’s inertia with drag and gravity and is 
\begin{equation}
    \frac{d\tilde{\boldsymbol{u}}}{d\tilde{t}} = \tilde{\boldsymbol{f}}_{d,p} + \tilde{\boldsymbol{g}}, 
\label{particle dynamics}
\end{equation}
where $\tilde{\boldsymbol{u}}$ is the particle velocity, 
tildes denote quantities with units, and the coordinate system is in Fig.~\ref{coordinates}. 
Additional terms are needed to capture nonzero Reynolds-number and fluid-inertia effects, 
which we neglect since the dynamics produced by Eq.~\ref{particle dynamics} captures the inertial-particle phenomena of interest here \cite[][]{MaxeyparticleEQOM}. 

Drag on small particles is linear in the velocity relative to the fluid, 
and the specific drag force is 
\begin{equation}
    \tilde{\boldsymbol{f}}_{d,p} = (\tilde{\boldsymbol{w}}-\tilde{\boldsymbol{u}})/\tau_p, 
    \label{particle force}
\end{equation}
where $\tau_p$ is the characteristic response time of the particle and is large for massive, inertial particles. 
For particles at low Reynolds numbers, $\tau_p$ is given by Stokes’ law, $\tau_p = \rho d^2/18\mu$, where $\rho$ and $\mu$ are the density and viscosity of the fluid, and $d$ is the diameter of the particle \cite[e.g.][]{WangandMaxey}. 
The fluctuating fluid velocity in the vicinity of the particle is $\tilde{\boldsymbol{w}}$, 
which is not modified by the presence of the particle in this model,
and is given by measurements or by solutions to the Navier-Stokes equations for the fluid. 
We let $\tilde{\boldsymbol{g}} = -\tilde{g}\hat{\boldsymbol{e}}_3$, 
as in Fig.~\ref{coordinates}, 
and we do not model the particle orientation \cite[][]{MaxeyparticleEQOM}. 

We make Eq.~\ref{particle dynamics} dimensionless with the characteristic velocity and length scales of the turbulence, $U$ and $L$, respectively, and incorporate Eq.~\ref{particle force} so that 
\begin{equation}
    \frac{d\boldsymbol{u}}{dt} = \frac{1}{St_{p}}(\boldsymbol{w} - \boldsymbol{u} - W_p\hat{\boldsymbol{e}}_3), 
    \label{dimensionless particle dynamics}
\end{equation}
where the Stokes number, $St_p = \tau_pU/L$, compares the characteristic turbulence and particle timescales and is large for heavy particles, and the settling parameter $W_p = U_{QF,p}/U$ is the ratio of the particle’s settling velocity through quiescent fluid, $U_{QF,p} = \tau_p g$, to the characteristic velocity of the turbulence. 
In general, the perturbations caused by turbulence lead to increased path lengths for particles settling through the fluid. Intuition may suggest, then, that settling times generally increase through turbulent fluid relative to quiescent fluid, but this is not the case. 

An interesting feature of solutions to Eq.~\ref{dimensionless particle dynamics} is that the mean particle velocity (in the direction of ${\boldsymbol{g}}$), is larger in a turbulent flow than in a quiescent flow \cite[][]{Maxey87}. 
The surface of mean settling velocity, which depends on $St_p$ and $W_p$, has a basin of increased velocity as its single feature of interest. 
This basin is centered near normalized particle inertia and velocity of order one. 
The phenomenon, called fast tracking \cite[][]{MaxeyCorrsin86}, 
occurs despite path lengths being increased by turbulence. 
An eddy moving opposite a particle’s direction of motion tends to push the particle away, causing the particle to move into a new eddy. 
On the other hand, eddies with the same direction of motion as the particle sweep the particle along. 
In this way, particles tend to to be swept into those parts of a turbulent flow with tailwinds without need for sensors or computation. 

In the following sections we define and characterize a cyber-physical forcing designed to produce fast tracking in flight vehicles even if a vehicle’s inertia and airspeed are not appropriately tuned with the flow in the way that produces fast-tracking in particles.

\subsection{Flight vehicle dynamics (DN)}

In order to generate qualitative insight, we treat flight vehicles theoretically like small particles characterized only by their mass, by a drag force proportional to their motion relative to air, and by a body force. 
For small particles, the body force is gravity, while for flight vehicles it is the thrust that keeps them aloft and propels them toward a given destination. 
While this model ignores many important aspects of flight vehicle dynamics \cite[e.g.][]{helicopter-theory}, it is commonly used for rotorcraft and fixed-wing flight control problems both with and without turbulence \cite[e.g.][]{droneswarm,crazyflypointmass,Patel2009ExtractingEF}, and explains some observed behaviors of birds flying through the turbulent atmosphere \cite[e.g.][]{Kaseybirdturbulence}. 
Our flight-vehicle momentum equation is then 
\begin{equation}
    \frac{d\tilde{\boldsymbol{u}}}{d\tilde{t}} = \tilde{\boldsymbol{f}}_d + \tilde{\boldsymbol{g}} + \tilde{\boldsymbol{f}}_T + \tilde{\boldsymbol{f}}_C. 
    \label{quadcopter dynamics}
\end{equation}
We explain the various terms in the following paragraphs. 
Drag is linear in the relative velocity for small particles (Eq.~\ref{particle force}). 
Though drag is generically quadratic, and not linear, 
for macroscopic flight vehicles at large Reynolds numbers \cite[][]{helicopter-theory}, 
we Taylor approximate the drag about its mean to first order. 
\begin{equation}
    \tilde{\boldsymbol{f}}_{d} = (\tilde{\boldsymbol{w}}-\tilde{\boldsymbol{u}} + \frac{1}{2} U_{QF} \hat{\boldsymbol{e}}_2)/\tau_d, 
    \label{quaddrag}
\end{equation}
which holds for small perturbations around an airspeed, $U_{QF}$, determined by the thrust defined below, and by the time constant, $\tau_d$, that characterizes the response of the flight vehicle to changes in airspeed. 
Note that fully nonlinear drag can cause loitering, the opposite of fast tracking \cite[][]{Good}, but that flight can nonetheless be enhanced beyond the baseline set by nonlinear drag with the cyber-physical methods introduced here. 
The form of the drag does not change our qualitative conclusions, and arbitrary nonlinearity can be incorporated into the flight vehicle model by modifying Eq.~\ref{quaddrag}. 

We let the specific thrust, $\tilde{\boldsymbol{f}}_T$, have one component that balances gravity so that the vehicle maintains altitude, and another component that maintains a certain airspeed, $U_{QF}$, through quiescent fluid given by $\tilde{f}_0 = 3 U_{QF}/2\tau_d$, so that
\begin{equation}
    \tilde{\boldsymbol{f}}_T = g \hat{\boldsymbol{e}}_3 - \tilde{f}_0 \hat{\boldsymbol{e}}_2.
\end{equation}
Physically, $\tilde{f}_0$ constantly pushes the flight vehicle toward its destination, which is at infinity in the $-\hat{\boldsymbol{e}}_2$ direction, and which in practice requires that the vehicle know its orientation and that it keep a fixed component of its thrust pointed toward the destination with an orientation controller that is not part of our analysis. 
In other words, we assume that rotational degrees of freedom were controlled quickly enough to produce desired translations, which is justified by the separation in scales between the integral length scales of atmospheric turbulence and the size and response time of most rotorcraft. 
An additional thrust force, 
$\tilde{\boldsymbol{f}}_C$, is unconstrained in general except by requirements on the stability and performance of the flight vehicle, which are beyond the scope of this study. 
We introduce a specific form for this forcing in the next section.

\subsection{Cyber-physical flight vehicle dynamics (FT)}
Here we summarize the selection of a particular forcing 
and of particular values for its free parameters. 
We show under certain conditions that the governing equation for a flight vehicle is the same as the one for a falling particle, though in a horizontal rather than vertical plane. 
This means that the inertial particle literature can be applied to the analysis of fast-tracking flight vehicles. 
To change the vehicle's dynamics under the constraint that it mimic particle dynamics, the forcing, $\tilde{\boldsymbol{f}}_C$, could imitate either particle inertia or drag. 
We choose to generate an effective inertia different from the vehicle’s real inertia with a force proportional to acceleration, $\tilde{\boldsymbol{f}}_C = C\,d\tilde{\boldsymbol{u}}/d\tilde{t}$, where $C$ is a dimensionless constant that we call the virtual inertia. 
Real inertia is isotropic and positive definite. 
Virtual inertia in contrast can be positive or negative, as well as anisotropic. 
As a result it can increase or reduce the effective inertia of a flight vehicle, which is the sum of its real and virtual inertias. 
That is, the virtual inertia can be adjusted to make a lightweight vehicle act like a heavier one, for instance. 
The only measurements needed to implement the forcing are given by on-board accelerometers 
-- the flight vehicle itself is the only probe necessary and no flow measurements are needed. 

We introduce anisotropy in the virtual vehicle inertia as an archetypal modification to particle physics that might extend the advantages of fast-tracking to more vehicles and conditions. 
To do so, we let 
\begin{equation}
    \tilde{\boldsymbol{f}}_{C} = \mathsfbi{C} \,  \frac{d\tilde{\boldsymbol{u}}}{d\tilde{t}}, 
\end{equation}
where $\tilde{\boldsymbol{f}}_{C}$ is a vector and $\mathsfbi{C}$ is a 2$\times$2 matrix. 
We consider only diagonal matrices of the form 
\begin{equation}
    \mathsfbi{C} = \begin{bmatrix} c_1 & 0 \\ 0 & c_2 \\ \end{bmatrix}, 
    \label{mass matrix}
\end{equation}
where $c_1$ and $c_2$ are dimensionless virtual masses. 
When they are larger than zero, they reduce the effective inertia of the flight vehicle in the horizontal plane. 
When they approach one, it is as if the vehicle inertia disappears asymptotically and the vehicle velocity approaches the fluid velocity as explained below. 

Finally, we combine Eqs.~\ref{quadcopter dynamics} through \ref{mass matrix}, 
and make the resulting equation dimensionless with characteristic velocity and length scales of the turbulence, $U$ and $L$, respectively. 
In terms of dimensionless variables, which do not have a tilde, the result is 
\begin{equation}
    \frac{d\boldsymbol{u}}{dt} = \frac{1}{M \, St}\begin{bmatrix} 1 & 0 \\ 0 & 1/A \\ \end{bmatrix} (\boldsymbol{w} - \boldsymbol{u} - W\hat{\boldsymbol{e}}_2),
    \label{FT Dynamics}
\end{equation}
The number $M = 1 - c_1$ is the factor by which the effective inertia of the flight vehicle is different from its actual inertia, and is larger than one for vehicles that act as if they had more inertia than they really do in the horizontal direction perpendicular to the average flight direction. 
The factor $A = (1-c_2)/(1-c_1)$ is the anisotropy in the effective inertia and is larger than one for vehicles that have more effective inertia in the direction of flight than perpendicular to it. 
Finally, $W = U_{QF}/U$ is the ratio of the flight vehicle’s speed through quiescent fluid to the characteristic velocity of the turbulence, and gravity does not contribute to the dynamics since it has been canceled by one component of the thrust. 

The solutions to Eq.~\ref{FT Dynamics} depend on three dimensionless quantities, $M \, St$, $A$, and $W$. 
Flight vehicles for which $M \, St$ is small respond more quickly than $\boldsymbol{w}(t)$ changes in time, in which case Eq.~\ref{FT Dynamics} can be integrated to show that the vehicle's velocity, $\boldsymbol{u}(t)$, relaxes exponentially to $\boldsymbol{w} - W\hat{\boldsymbol{e}}_2$ at a rate determined by $M \, St$. 
When $M$ or $St$ approach zero the vehicle loses its inertia and it moves with the flow; when $M$ is negative the vehicle's velocity diverges from the flow velocity exponentially and the flight is unstable. 

For isotropic flight vehicles, for which $A$ is equal to one, Eq.~\ref{FT Dynamics} is identical to the one for a particle settling through turbulence under gravity (Eq.~\ref{dimensionless particle dynamics}) with the parameter $M \, St$ taking the place of $St_p$, and the $\tilde{f}_0$ component of the thrust playing the role of gravity in the definition of $W$. 
Up to differences introduced by anisotropy in the virtual inertia, fast-tracking is therefore a feature of flight vehicle dynamics as it is for particles. 
The question we next address is what values of $\tilde{f}_0$, $c_1$, and $c_2$ are useful to achieve certain objectives, which we do in terms of their dimensionless representatives $W$, $M \, St$, and $A$.

\subsection{Cyber-physical flight vehicle power requirements}
\label{Sec:CPPR}
The dynamic model of an isotropic flight vehicle in Eq.~\ref{FT Dynamics} is identical to the one for a falling particle, Eq.~\ref{dimensionless particle dynamics}, but the energetics of each are different. 
A particle exchanges potential energy with kinetic energy and drag, while a flight vehicle expends energy to produce thrust both to stay aloft and to generate other desired motions. 
We constrain the coefficients, $A$ and $M$, of the forcing in Eq.~\ref{FT Dynamics} either by minimizing the energy required for flight or by maximizing average speed for a given energy. 
We next estimate the work performed by the forcing to generate the desired motions and deviations from unforced flight trajectories. 

To derive the energy equation we consider rotorcraft that automatically rotate to point their propeller axes into the direction of the net thrust, 
and for which the power required can be determined from functions of the $l^2$ norm of the net thrust, $\boldsymbol{F}_T$, alone. 
The approximate power, $\tilde{P}$, is 
\begin{equation}
    \tilde{P} = {c}_P (\tilde{\boldsymbol{F}}_T^2)^{n}, 
    \label{rotor power}
\end{equation}
where $n = 3/4$ in the limit of large induced flow and small propeller advance ratio according to actuator disk theory \cite[][]{helicopter-theory}, but could take other values. 
The coefficient, ${c}_P$, has the dimensions of  $\tilde{P}/\tilde{F}^{2n}$ and depends on the fluid density, propeller geometry, and efficiency. 
Since we sought scaling laws and ${c}_P$ is a constant, we do not specify it. 
Depending on flight speed, the expression for $\tilde{P}$ is more complex than Eq.~\ref{rotor power} \cite[][]{helicopter-theory}. 
However, only the local curvature of $\tilde{P}$ is important in our analysis since we considered small changes in thrust, and any local curvature in $\tilde{P}$ can be modeled by adjusting $n$. 
We found that our results did not change qualitatively when $n$ was varied about $n = 3/4$ within physical bounds. 

To compute the power, we recombine the components of the thrust, 
which we until now had split into parts, so that 
\begin{equation}
    \tilde{\boldsymbol{F}}_T 
        = m(\tilde{\boldsymbol{f}}_T + \tilde{\boldsymbol{f}}_C), 
\end{equation}
where $m$ is the mass of the flight vehicle. 
The dimensionless power, $P = \tilde{P} / c_P (mg)^{2n}$, is then composed of four parts, two resulting from the work performed to accelerate the vehicle in the plane of motion, one from the constant thrust toward the destination, $-{\tilde{f}}_0\hat{\boldsymbol{e}}_2$, and one from the work against gravity. 
We regroup these terms according to the dimensionless variables identified above and a new one called $G = g\tau_d / U$, which normalizes the (inverse) strength of turbulence, so that 
\begin{equation}
    P = \left[\left(\frac{St}{G}(1 - M) \frac{du_1}{dt} \right)^2 + \left( \frac{St}{G}(1 - M\,A) \frac{du_2}{dt} - \frac{3}{2}\frac{W}{G}\right)^2 + 1\right]^n,
    \label{dimensionless power eqn}
\end{equation}
Observe that two dimensionless groups govern the power requirements for isotropic flight vehicles (when $A=1$), one being $(1 - M)St/G = (c_1/g)(U^2/L)$, which is the flow-normalized virtual mass-to-weight ratio, or the virtual Stokes number to turbulence-intensity ratio, and the other being $W/G = 2\tilde{f}_0/3g$, which is a thrust-to-weight ratio. 
In other words, the dimensionless variables that govern the energetics are different from those that govern the dynamics, with flow properties providing natural units for the dynamics and the vehicle’s weight doing so for the energetics. 
The parameters $St$ and $G$ form an independent pair that fully characterize the flow and flight vehicle irrespective of the forcing, and we used this pair, rather than their combinations with $W$, $M$ and $A$ in the power equation, to describe the system configuration. 

The action of the forcing is embodied in the variables $W$, $M$ and $A$. 
For unmodified inertia, when the latter two variables are equal to one, the power required for flight is determined only by the thrust-to-weight ratio, $P\sim 1 + 9W^2/4G^2$, and not by the accelerations of the vehicle. 
Note that $W/G$ can be interpreted as the tangent of the rotorcraft’s equilibrium angle of lean during flight through quiescent fluid, and represents how hard the rotorcraft works to stay aloft relative to how hard it works to move forward. 
For hovering vehicles, $W/G=0$, while fast flight on a planet with weak gravity corresponds to large $W/G$. 
Finally, the power required by neutrally-buoyant vehicles can be modeled roughly by Eq.~\ref{dimensionless power eqn} without the +1, though our dynamical equation, Eq.~\ref{quadcopter dynamics}, would then also need to incorporate terms that capture the effects of fluid inertia, 
which we neglected for simplicity since they do not change our qualitative conclusions.

\subsection{Cyber-physical flight vehicle energy approximation}
Our objective is to find sets of parameters for the forcing that cause the flight vehicle to fast-track, or to reach a certain destination with a net benefit either in energy or time expended. 
Therefore, we are interested only in low-energy solutions, or only in those sets of controlled parameters that govern power, $W$, $M \, St$, and $A$, for which the energy needed to fast track does not exceed the energy gained by doing so. 
The energy being the time integral of the power given by Eq.~\ref{dimensionless power eqn}, observe that the only time-dependent terms are the ones proportional to the accelerations, $du_i/dt$, and that these terms are mixed with others under exponents. 
In order to isolate the time dependence and so to facilitate integration, we expand around small values of the time-dependent terms, recognizing that these small values correspond to the low-energy solutions of interest. 
Other choices for expansions lead to similar results. 
We find that not only is the energy easier to calculate, but that it depends on only one statistic of the vehicle’s trajectory, which is the variance of its accelerations, and not on any other property of the trajectory. 

The efficiency of transportation vehicles can be measured by 
the cost of transport, $E$ \cite[][]{pricespeed}. 
It is the time integral of power 
per unit weight and unit distance traveled, 
$E = \tilde{E} / (mg \tilde{d})$, 
where $\tilde{E}$ is the energy 
required to travel a given distance $\tilde{d}$ 
in the $-\hat{\boldsymbol{e}}_2$ direction. 
Since the cost of transport is proportional to energy, 
and since we specify $mg$ and $\tilde{d}$ {\it a priori}, 
we refer to the cost of transport succinctly as ``energy'' 
or ``dimensionless energy'' throughout the paper. 
The dimensionless energy is then 
\begin{equation}
    E = 
        \frac{1}{mg \tilde{d}} 
            \int_0^{\tilde{t}_f} \! \tilde{P} \, d\tilde{t}, 
    \label{eq:newCOT}
\end{equation}
where $\tilde{t}_f$ is the time 
required to travel the distance $\tilde{d}$. 
Note 
that only the integrand and limit of integration are flow-dependent, 
and not the prefactors. 
We rewrite the right-hand-side of Eq.~\ref{eq:newCOT} in dimensionless variables, 
so that 
\begin{equation}
    E = C_P \frac{G}{d} \int_{0}^{t_f} \! P \, dt, 
    \label{energy integral}
\end{equation}
where $C_P = (mg)^{2n-1} / (g \tau_d)$ is constant, 
and $t_f = \tilde{t}_f U/L$ 
and $d = \tilde{d}/L$ 
are the number of flow time and length scales traveled by the rotorcraft, respectively. 
In quiescent flow, where $L$ is undefined, 
$L$ is an arbitrary reference length, 
and the flow timescale $L/U$ cancels out upon integration of the (constant) power. 

The power required by the flight vehicle is determined by the accelerations it experiences, 
which are functions of time, 
position and the parameters that govern the dynamics, 
so that we can rewrite the dimensionless power equation (Eq.~\ref{dimensionless power eqn}) 
in terms of two functions, $\mathrm{f}_1$ and $\mathrm{f}_2$, as 
\begin{equation}
    P = \left(\mathrm{f}_1^2(t, \boldsymbol{u}, W, M\,St, A) 
        + (\mathrm{f}_2(t, \boldsymbol{u}, W, M\,St, A) - 3W/2G)^2 
        + 1\right)^n, 
    \label{pform}
\end{equation}
where $P=P(\mathrm{f}_1,\mathrm{f}_2)$ is a functional that we expanded in a Maclaurin series 
for small $\mathrm{f}_1$ and $\mathrm{f}_2$. 
At the second order, we have 
\begin{multline}
    P(\Delta \mathrm{f}_1, \Delta \mathrm{f}_2) = 
        P(0,0) 
        + \mathrm{f}_1\frac{\partial P}{\partial \mathrm{f}_1} \Bigr\rvert_{0,0} 
        + \mathrm{f}_2\frac{\partial P}{\partial \mathrm{f}_2} \Bigr\rvert_{0,0} \\
        + \frac{1}{2} \left(\mathrm{f}_1^2 
            \frac{\partial^2 P}{\partial \mathrm{f}_1^2} \Big\rvert_{0,0} 
        + \mathrm{f}_2^2 \frac{\partial^2 P}{\partial \mathrm{f}_2^2} \Big\rvert_{0,0} 
        + 2\mathrm{f}_1\mathrm{f}_2 \frac{\partial^2 P}{\partial \mathrm{f}_1 \partial \mathrm{f}_2} 
            \Big\rvert_{0,0} \right)
        + O(\mathrm{f}_i^3), 
    \label{Mclaurin expansion}
\end{multline}
where the mixed partial derivative is zero given the form of Eq.~\ref{pform}. 

We simplify the expression for $E$ (Eq.~\ref{energy integral}), 
which is exact, 
with the expansion in Eq.~\ref{Mclaurin expansion}, 
and find that the approximation, 
\begin{equation}
    E \, \approx \,C_P T \left(\$_G+\$_1 + \$_2\right),
\label{approximate energy}
\end{equation}
holds under certain conditions discussed below, 
where $T \equiv t_f G/d = (\tilde{t_f}/\tilde{d})g\tau_d$ 
normalizes average ground speed ($\tilde{d}/\tilde{t_f}$), 
which is variable, 
by a gravitational velocity scale for the rotorcraft ($g\tau_d$), 
which is constant. 
The expansion simplifies the expression for energy 
since the integrals in Eq.~\ref{energy integral} 
become moments of acceleration statistics. 
This can be seen in the energetic costs, 
which are given by 
\begin{equation}
    \$_G = \left( 1 + \frac{9}{4} \frac{W^2}{G^2} \right)^n, 
    \label{Flight Power Components}
\end{equation}
\[
    \$_1 = n \$_G^{1-\frac{1}{n}} \frac{St^2}{G^2} (1-M)^2 
            \alpha_1, \,\, \mbox{and}
\]
\[
    \$_2 = n \$_G^{1-\frac{2}{n}} 
            \left( (2n-1) \frac{9}{4} \frac{W^2}{G^2} + 1 \right) 
            \frac{St^2}{G^2} (1-MA)^2 \alpha_2, 
\]
where $\alpha_1$ and $\alpha_2$ are the variances of the accelerations 
experienced by the flight vehicle, 
\begin{equation}
    \alpha_i(W, M\,St, A) 
        = \frac{1}{t_{f}} 
        \int_{0}^{t_{f}} \! \left( \frac{du_i}{dt} \right)^2 dt, 
\end{equation}
and where $i$ is either 2 or 1, the direction of flight or orthogonal to it, respectively. 
The integrals of the linear terms in Eq.~\ref{energy integral} are approximately zero according to the fundamental theorem of calculus, since the expectation value for the difference between initial and final velocities is zero over many independent realizations of turbulence. 
As a result, those terms do not appear in Eq.~\ref{Flight Power Components}. 

We comment briefly on the higher-order terms in the expansion (Eq.~\ref{Mclaurin expansion}). 
Once integrated to obtain energy, they are proportional to increasing powers, $m$, of the acceleration variance multiplied by the moments of the acceleration distribution, 
$M_m \equiv \langle (du_i/dt)^m \rangle / \langle (du_i/dt)^2 \rangle^{1/m}$. 
The tail of the distribution of inertial particle accelerations is bounded from above by the distribution of fluid particle accelerations 
\cite[][]{Warhaftparticleacceleration}, 
which can be described empirically by a stretched exponential 
\cite[e.g.][]{fluidAccelerationPdf}, 
and whose corresponding moments depend on the Reynolds number of the turbulence \cite[e.g.][]{PortaHighReAcceleration}. 
The expansion therefore holds 
to the extent that $\langle \mathrm{f}_1^2 \rangle < 1$ and $\langle \mathrm{f}_2^2 \rangle < 1$, both of which are proportional to the acceleration variance, and that the moments converge for increasing $m$ and Reynolds numbers. 
Note that by controlling the size of $\langle \mathrm{f}_1^2\rangle$ and $\langle \mathrm{f}_2^2\rangle$ (by changing $c_1$ and $c_2$ for instance), the energy approximation can be made arbitrarily accurate on any time interval $t\in (t_a,t_b)$ for which $|d\boldsymbol{u}/dt|<\infty$.

The expansion ignores changes in sign of $\mathrm{f}_2 - 3W/2G$, 
which are likely to occur under intermittent large accelerations. 
Therefore the expansion underestimates energy consumption in principle 
-- the energy equation is valid only when the forcing does not push backward harder than does the specific thrust in the forward direction. 
By comparing terms in the following way, we find that this effect is negligible 
except perhaps for flight vehicles with a lower $G$ and $St$ than any we investigated. 
If $\$_G \approx 3^n$ (see Sec.~\ref{benchmark thrust}), 
this implies $\$_1 = 3^{n-1} n \langle \mathrm{f}_1^2 \rangle$. 
If in addition, $n > 1/2$ then $\$_2 = 3^{n-2} n (4n-1) \langle \mathrm{f}_2^2 \rangle$. 
We verified the approximation by comparing the average power components $\$_1$, 
and $\$_2$, to the components $\$_G$ and 1, 
a comparison that was favorable. 
Furthermore, 
we did not observe in our calculations any instantaneous extreme accelerations 
that reversed the sign of the term in question, 
but such extreme events may be more likely in real turbulence than in our model turbulence and the matter is worth future investigation. 

For any given set of dimensionless parameters, evaluation of the energy equation (Eq.~\ref{approximate energy}) requires computer simulations to determine $T$, $\alpha_1$, and $\alpha_2$. Since each of these variables is determined by the system's dynamics, each is then a smooth scalar function of $W$, $M \, St$, and $A$. Therefore, $T$, $\alpha_1$, and $\alpha_2$ are described by three-dimensional manifolds embedded in four dimensions. 
When referring to these manifolds, we identify a particular point on them by $W$, $M \, St$, and $A$ such that the unique point on the respective manifold with those coordinates is $T$, $\alpha_1$, or $\alpha_2$. For example, when referring to the minimum of the time manifold, the manifold for $T$, we are considering the point $(W, M \, St, A)$ such that $T$ achieves its minimum value on the manifold. It is not relevant to our problem to sample from the manifold in other coordinate systems. These manifolds need to be estimated stochastically for given turbulent velocity fields. 

All terms within the parentheses of Eq.~\ref{approximate energy} represent costs, with the first, $\$_G$, being the (constant) power required to stay aloft plus the power used to produce thrust toward the destination. The two terms proportional to accelerations, $\$_1$ and $\$_2$, are the average power used to produce the forcing, and are zero either if it is switched off or when flying through quiescent fluid. The costs diverge toward infinity for small $G$. Net energetic benefits are realized by a reduction in flight time, $T$. Various limiting cases indicate the relative importance of terms, suggest universal functional dependencies, and point to applications where the control ideas, if they work, would be useful. One such limit establishes a certain optimized thrust, discussed in the next section.


\subsection{Benchmark thrust}
\label{benchmark thrust}
As a reference, we calculate the optimum airspeed (or thrust) in the absence of turbulence for which energy is minimized. In the absence of turbulence, and therefore of accelerations so that $\alpha_1$ and $\alpha_2$ are zero, the energy required to move according to Eq.~\ref{approximate energy} is given by 
\begin{equation}
    E_{QF} = C_PT_{QF} \$_G. 
    \label{QF energy}
\end{equation}
Since the dimensionless transit time across an arbitrary distance through quiescent fluid, $T_{QF} = (L/U_{QF})(g\tau_d/L) = G/W$, is given by the inverse of the velocity, the energy in Eq.~\ref{QF energy} can be re-expressed exactly as 
\begin{equation}
    E_{QF} = C_P \frac{G}{W} 
                \left( 1 + \frac{9}{4} \frac{W^2}{G^2} \right)^n. 
    \label{benchmark energy}
\end{equation}
Energy is minimized for a particular value of the thrust-to-weight ratio, namely 
\begin{equation}
    \frac{W^*}{G^*} = \frac{2}{3} \sqrt{\frac{1}{2n-1}}, 
\end{equation}
which is equal to $(2/3)\sqrt{2}$ when $n = 3/4$. 
In other words, in a quiescent fluid and given the set of parameters that describe the flight vehicle, the mean thrust to minimize energy consumption has an optimal value, for which the corresponding airspeed is 
$U_0^*=\tilde{f}^*_0\tau_d=(2/3)(2n-1)^{-1/2}g\tau_d$. 
If the optimum thrust were maintained in turbulence, airspeed would be perturbed but would continuously relax exponentially to $U_0^*$ according to the dynamics in Eq.~\ref{quadcopter dynamics}. 
We therefore use these optimum values for thrust (and airspeed) to evaluate the do-nothing (DN) dynamics determined by Eq.~\ref{quadcopter dynamics}, and as benchmarks against which to compare improvements made by FT forcing. 
One main conclusion is that turbulence moves the optimal $W/G$ away from $W^*/G^*$ under many conditions.

\subsection{Disturbance Rejection (DR)}
One way to respond to disturbances caused by turbulence to a flight trajectory is to reject them and so to maintain an approximately straight trajectory. 
Within the context of the models presented above, we evaluate the work required to fly straight as the one developed by an isotropic forcing with infinite virtual inertia, for which $A = 1$ and $M \to \infty$. In this way, we can evaluate the energetic cost of disturbance rejection. 

\begin{figure}
  \centerline{\includegraphics[width=0.5\linewidth]{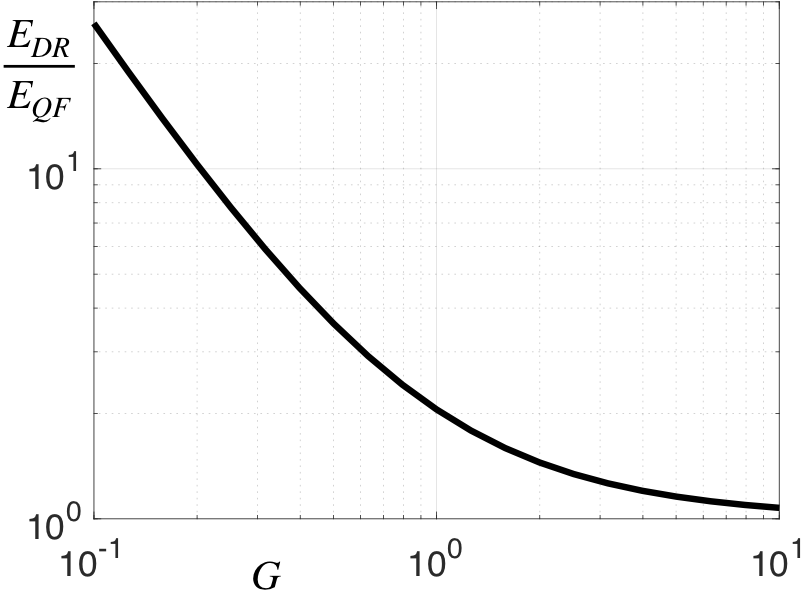}}
  \caption{The energetic cost of disturbance rejection ($E_{DR}$) is always larger than the cost of flight through quiescent fluid ($E_{QF}$) under environmental conditions given by $G$ according to Eq.~\ref{DR energy}. 
}
\label{DR plot}
\end{figure}

As the mass multiplier, $M$, diverges to infinity, the accelerations experienced by a flight vehicle approach zero, so that the costs in Eq.~\ref{Flight Power Components} look at first indeterminate. From Eq.~\ref{FT Dynamics}, observe that the vehicle’s accelerations are inversely proportional to $M \, St$, so that the acceleration statistics scale in the same way. The mean-square accelerations, $\alpha_1$ and $\alpha_2$, then scale with the inverse of $M^2 St^2$. The costs, when ignoring the accelerations, are explicitly proportional to $M^2 St^2$ for large $M$, which cancels out the scaling of the accelerations. For $A=1$ and $M \to \infty$, the costs therefore approach constants determined by $W$, $G$, and $c$, where $c$ is a proportionality constant that needs to be determined empirically. We substitute these constants back into the energy equation, Eq.~\ref{approximate energy}, and use the benchmark thrust defined above, $W^*/G^*$, to find that 
\begin{equation}
    \frac{E_{DR}}{E_{QF}} 
        = \left( \frac{2n-1}{2n} \left( \frac{c^2}{G^2} 
        + \left( \frac{c}{G} 
        + \sqrt{ \frac{1}{2n-1}} \right)^2 + 1 \right) \right)^{-n}. 
    \label{DR energy}
\end{equation}
The energetic cost of disturbance rejection 
diverges toward infinity for increasing turbulence intensity, and only approaches one (from above) for vanishing turbulence. 
As seen in Fig.~\ref{DR plot} 
for $c^2\approx0.5$, which we determined empirically, and $n=3/4$, 
not only is DR never energetically favorable, 
but working against turbulence also eliminates fast-tracking and its advantages.
Simply relaxing DR would be beneficial 
if it were possible to do so while maintaining stability, 
which is a problem that is beyond the scope of this study. 

\subsection{Parameter space mapping}
We treat the optimization process as a mapping from each set of given parameters, $G$ and $St$, to a set of dynamic parameters $W, M \, St$ and $A$ that minimized energy or flight time. 
In this sense, the forcing (FT) is simply a vector valued function, or mapping, whose inputs are $G$ and $St$, and whose outputs are $W, M \, St$ and $A$.  
The purpose of optimization is to find this function. 
The DN and DR strategies are also vector valued functions of the input variables $G$ and $St$, however these functions are not guaranteed to, and indeed rarely did, output $W, M \, St$ and $A$ that minimized either energy or flight time for a given energy. 
This mapping viewpoint is useful because it yields physical insight. 

First we define a mapping from the set of dimensionless parameters that are constrained by the characteristics of the turbulence and flight vehicle, $G=\tau_dg/U$ and $St=\tau_dU/L$, into the set of dimensionless parameters that are freely adjustable during optimization and that govern the dynamics, $W=U_0/U$, $M \, St = (1-c_1)\tau_dU/L$, and $A=(1-c_2)/(1-c_1)$. 
We start with the mapping for the FT forcing, which can be defined as a vector field in three dimensions as follows: 
\[
G, St \in \mathbb{R}_+
\]
\[
W_{FT}(G,St), M_{FT}(G,St), A_{FT}(G,St): \mathbb{R}_+^2 \to \mathbb{R}_+
\]
\[
\boldsymbol{h}_{FT}: \mathbb{R}_+^2 \to \mathbb{R}_+^3
\]
\begin{equation}
    \boldsymbol{h}_{FT}(G,St) =
    \begin{bmatrix}
    W_{FT}(G,St)\\
    StM_{FT}(G,St)\\
    A_{FT}(G,St)
    \end{bmatrix},
\end{equation}
where $ \boldsymbol{h}$ is the mapping function between the constrained parameters, $G$ and $St$, and the free parameters $W$, $M \, St$, and $A$. We construct maps for the other strategies in the same way. 
Of particular importance is the DN strategy, 
\begin{equation}
    \boldsymbol{h}_{DN}(G,St)
=   \begin{bmatrix}
    \sqrt{1/(2n-1)}G\\
    St\\
    1
    \end{bmatrix}.
\end{equation}
Since $\boldsymbol{h}_{DN}$ is both injective and surjective with respect to input and dynamic parameters when $n>1/2$, we can compare FT and DN with a composite function using their respective mappings, $\boldsymbol{h}_{FT}$ and $\boldsymbol{h}_{DN}$. 
The composite mapping represents how much of the forcing used by an FT strategy is not already activated by the DN strategy, and is 
\begin{equation}
    \boldsymbol{J}\equiv \boldsymbol{h}_{FT}\left(\boldsymbol{h}_{DN}^{-1} \right).
    \label{composite mapping}
\end{equation}
Finally, we construct a vector field that contains a complete set of instructions for how to perform FT forcing for every set of input parameters, $(G,St)$, in the following way. 
In logarithmic space, the composite mapping in Eq.~\ref{composite mapping} is the ratio of the FT and DN controller’s authority, 
\begin{equation}
    \log{\boldsymbol{J}}=
\begin{bmatrix}
\log{W_{FT}(W_{DN}/\sqrt{1/(2n-1)},St)}\\
\log{StM_{FT}(W_{DN}/\sqrt{1/(2n-1)},St)}\\
\log{A_{FT}(W_{DN}/\sqrt{1/(2n-1)},St)}
\end{bmatrix}.
\end{equation}
This equation is a mapping from the na\"{i}ve parameters provided by the DN strategy to the set of parameters associated with an FT strategy, and is simplified by the fact that $M_{DN}=1$. 
We then construct a vector whose tail is located at the position given by the input to $\boldsymbol{J}$, $(\log{W_{DN}},\log{St_{DN}},0)$, and whose tip points to the FT parameters given by the output of $\boldsymbol{J}$. 
To simplify the presentation, we later show only the 2D projections of these mappings, and show the optimized values of $A_{FT}(G,St)$ in separate figures. 

If the time manifold, $T$, were constant everywhere, 
then the energy function’s Hessian would be positive definite with a minimum at $\log{\boldsymbol{J}} = \mathbf{0}$. 
As a result, the DN and FT strategies would be the same. 
However, if $T$ is not constant at the point $(\sqrt{1/(2n-1)}G,St,1)$, 
DN can no longer be optimal, and as a result $\boldsymbol{J}$ will be nonzero, at least local to those places where the gradient of $T$ is nonzero. 
This shows, even before performing computer simulations, that FT is likely to outperform DN 
since DN is contained within the space of possible FT strategies 
and cannot outperform FT, 
and the requirement that they perform equally is strict. 
These arguments do not indicate 
the extent to which FT outperforms DN, 
but indicate that a nonzero slope in $T$, rather than an offset of $T$, 
determines whether FT is beneficial. 
Furthermore, we see that even for the case that turbulence caused only loitering 
and not fast-tracking, FT would outperform DN.

\section{Methods}
\label{methodssection}
In this section we explain how we modeled turbulence in computer simulations, 
how we synthesized flight vehicle trajectories, 
and how we optimized FT.

\subsection{Flow simulation}
\label{sec:flowsimulation}

We use a two-dimensional (2D) implementation of the incompressible, statistically stationary, 
isotropic, and homogeneous turbulence model in \cite{Kraichnanturbulence} 
and employed to study fast tracking in \cite{Maxey87}. 
The model generates a power-law spectrum at low wavenumbers with an exponential cutoff at high wavenumbers, producing a peak in the spectrum and a flow with a single dominant length-scale. 
The model specifies the flow velocity, $\boldsymbol{w}$, according to a sum of random modes, 
\begin{equation}
    \boldsymbol{w} = \sum_{j=1}^N \boldsymbol{b}_n \cos{(\boldsymbol{k}_n\cdot\boldsymbol{x}+\omega_nt)}+\boldsymbol{c}_n \cos{(\boldsymbol{k}_n\cdot\boldsymbol{x}+\omega_nt)}, 
    \label{eq:w}
\end{equation}
where $\boldsymbol{x}$ is the position in the plane, 
and we used 64 modes as in \cite{Maxey87}. 
The parameters $\boldsymbol{b_n}$, $\boldsymbol{c_n}$, $\boldsymbol{k_n}$, and $\omega_n$ are drawn from a normal distribution, and $\boldsymbol{b_n}$ and $\boldsymbol{c_n}$ are subsequently conditioned to enforce incompressibility and the energy spectrum. 
Even though the flow is periodic, 
the periodicity occurs on an astronomical scale set 
by the lowest common multiple of the randomly chosen wavelengths \cite[][]{lcm:2019}, 
which constitutes an advantage of this random flow 
over direct numerical simulations of turbulence in periodic domains, 
since this domain is continually resampled by 
particles or flight vehicles in the high-speed limit. 

The turbulence model in Eq.~\ref{eq:w} 
is known to under-predict the strength of fast tracking for settling particles 
\cite[][]{WangandMaxey,Good}, 
but it predicts all the nontrivial qualitative behaviors needed for the investigation of fast-tracking energetics presented in this paper. 
For instance, 
the model incorporates the spatial and temporal correlations 
that are responsible for fast tracking -- 
uncorrelated flows cannot preferentially sweep particles, regardless of inertia. 
Note that particles can settle so quickly through turbulence 
that the flow changes more quickly than the particle can respond; 
for these particles the flow is effectively uncorrelated 
and the particles do not fast track. 
Furthermore, 2D flows must be time dependent in order to capture the behavior of particles with vanishing inertia. 
The path lines of these particles are the streamlines of an incompressible flow 
that is non-ergodic and periodic if the fluid flow is periodic. 
If $\boldsymbol{w}\cdot\hat{\boldsymbol{e}}_2>W_d$ 
then paths can form closed orbits and mean settling times for uniformly initialized particle ensembles become undefined. 
While particles with small but non-zero inertia follow the trajectories of a compressible flow \cite[][]{Maxey87}, their velocities are still uniquely specified by the flow and paths are periodic and non-ergodic if the underlying flow is also periodic \cite[][]{FalkovichCloud,BewleyCloud}. 
Finally, 
a continuum of scales is required in the flow to prevent strong loitering along a band of $St^2 W \sim 1$ \cite[][]{indefiniteloitering,vortexloitering}, and to reproduce the fundamentally multi-scaled nature of the fast tracking problem \cite[][]{Bragg2019}. The strong loitering band is unphysical in turbulence and corrupts the time manifold by becoming a dominant feature in it \cite[][]{MaxeyCorrsin86,Maxey87cell}. 
Motions on different scales allow flow structures to compete with each other 
and to disrupt loitering.

\subsection{Flight vehicle simulation}
Particle trajectories described by Eq.~\ref{FT Dynamics} 
were integrated using MATLAB’s {\it ode113} Adams-Bashforth-Moulton method. 
Each trajectory was integrated for a time of $t=4(AMSt+100(1+1/W))$, 
with only the last half being recorded. 
We found that this choice gave flight vehicle trajectories sufficient time to forget their initial conditions: the first term in the sum allowed all flow-independent initial conditions to be forgotten, while the second term allowed sufficient time for the flow-vehicle interactions to settle into dynamic equilibrium. A 15x15x13 grid of parameter values was tested for $10^{-5/2} \leq St \leq 10$, $0.1 \leq W \leq 10$, and $1 \leq A \leq 10^{10/3}$. At each grid point, vehicle trajectories were recorded for at least 20 randomly initialized flows. For each trajectory, settling time per unit distance was recorded as the inverse of settling speed. If the estimated settling time reduction was at least 5\%, simulations were run until the variance of estimated average settling time reduction was at most 20\% of the time reduction. This was a relative tolerance. If estimated settling time reduction was less than 5\%, simulations were run until the variance of estimated average settling time was less than 1\%. This was an absolute tolerance.

\begin{figure}
\centering
     \begin{subfigure}[b]{0.4\textwidth}
         \centering
         \includegraphics[width=\textwidth]{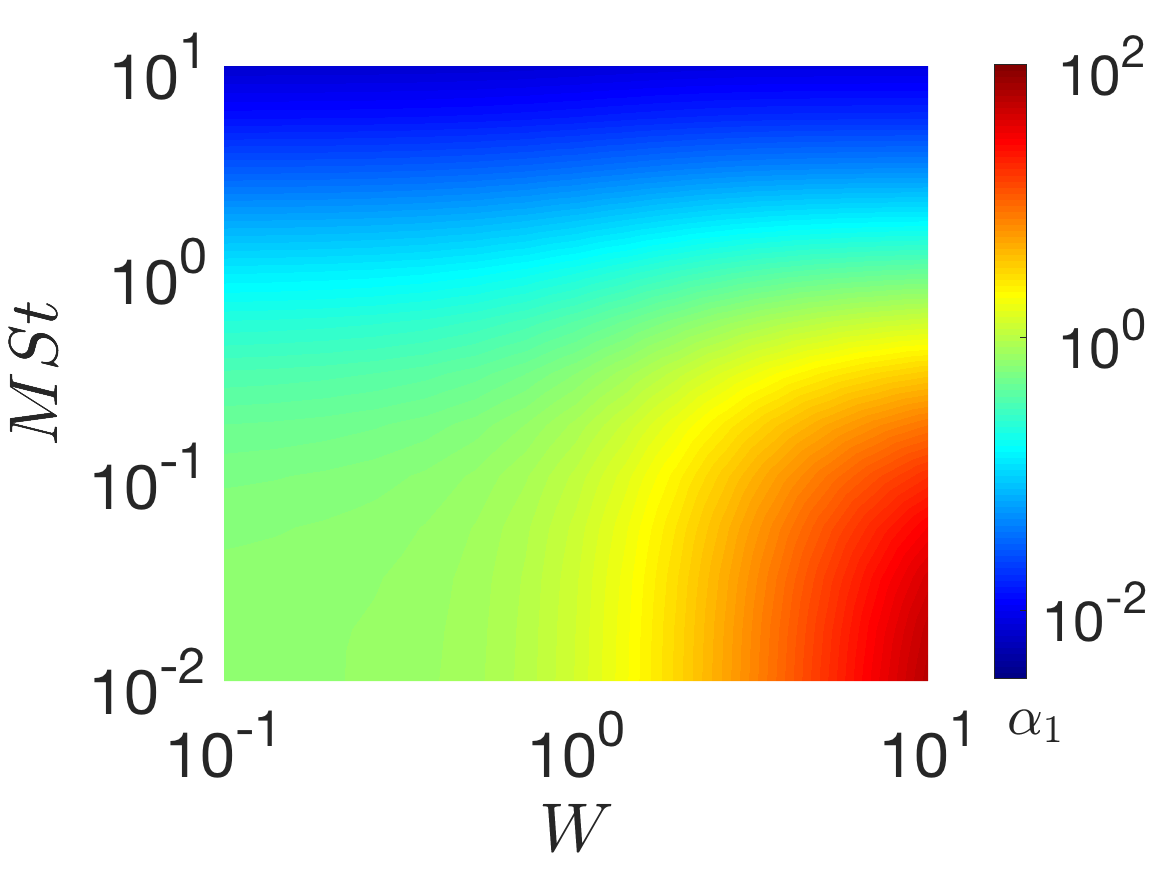}
         \label{acceleration 1}
     \end{subfigure}
     \begin{subfigure}[b]{0.4\textwidth}
         \centering
         \includegraphics[width=\textwidth]{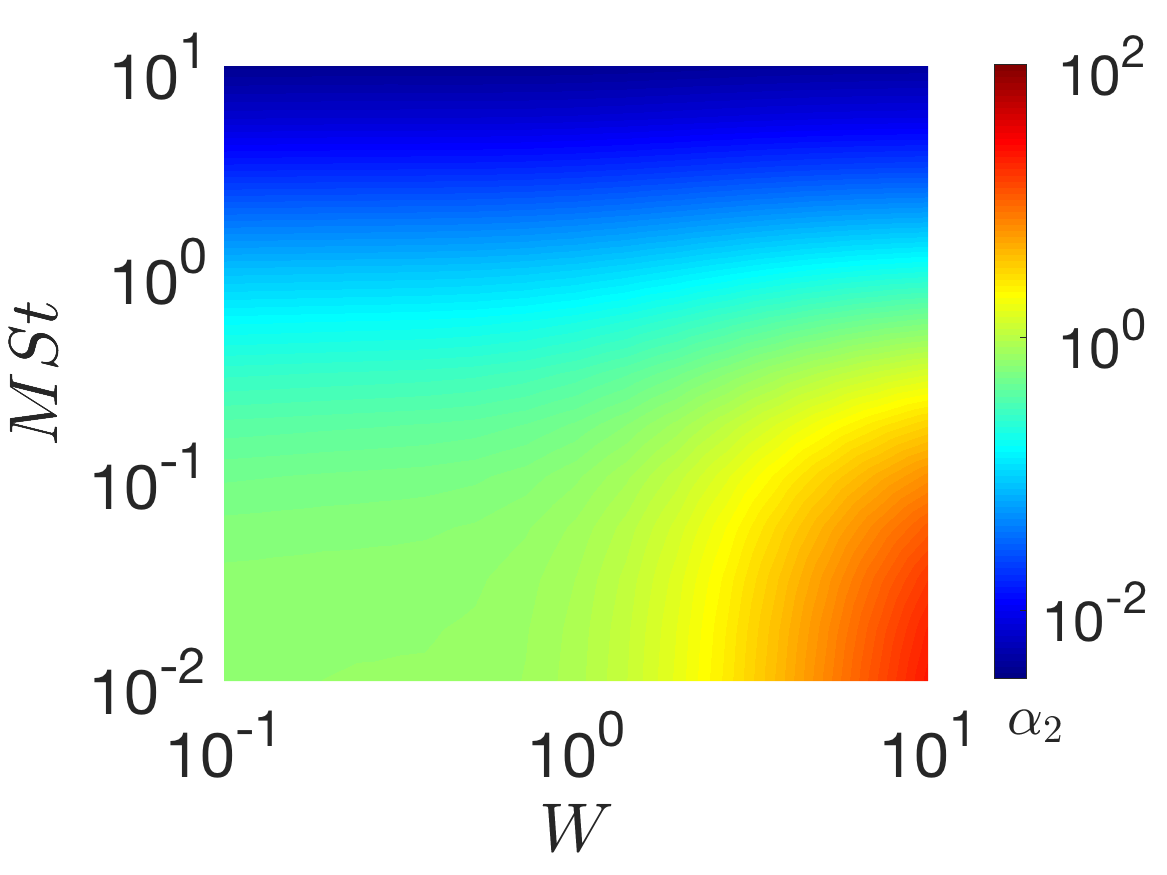}
         \label{acceleration 2}
     \end{subfigure}
\vspace{-0.2in}
  \caption{Acceleration variances for isotropic ($A$ = 1) particles and flight vehicles 
  traversing a turbulence model. 
  {\it Left:} The dimensionless variance of the accelerations, $\alpha_1$, 
  in the direction transverse to the one of mean flight. 
  {\it Right:} The accelerations, $\alpha_2$, in the direction of the flight's destination. 
  Similar manifolds describe motions for anisotropic ($A \neq$ 1) settling. 
  }
\label{alpha plot}
\end{figure}

Fig.~\ref{alpha plot} shows the smoothed acceleration variance, $\alpha_i$, for $A = 1$. 
It shows good qualitative agreement with previous studies on particle acceleration variance \cite[e.g.][]{Warhaftparticleacceleration}. 
For isotropic virtual inertia, the work required to mimic particle dynamics is proportional to these accelerations. 
Note that $\alpha_i$ is normalized by the characteristic flow velocity, and that it is $\alpha_i/W^2$ rather than $\alpha_i$ that properly reflects the real accelerations experienced by the flight vehicle. 
Particles or flight vehicles encounter large dimensionless accelerations when they are 
fast and lightweight (and large dimensional accelerations when they are slow and lightweight). 
The main effect of anisotropy is to shift $\alpha_2$ so as to suppress accelerations in the lower right-hand quadrant without significant modification to the shape of either manifold. 

\subsection{FT parameter optimization}
The two goals we consider are 
(1) to minimize energy consumption by all means available, and 
(2) to minimize transit time given a fixed energy budget, $E_{DN}$. 
The parameters we vary are the mean thrust toward the destination, captured in $W=U_0/U$, 
the effective inertia, captured in $M \, St = (1-c_1)\tau_dU/L$, 
and the anisotropy in the virtual inertia, captured in $A = (1-c_2)/(1-c_1)$. 
The parameters $G = \tau_dg/U$ and $St = \tau_dU/L$ represent the fixed environmental factors and were held fixed during optimizations. 
All case comparisons are made at a constant $G$ and $St$ except the no turbulence, 
quiescent flow (QF) case, for which $G$ is infinite. 
The FT optimization problem is therefore three dimensional. 
There are two cases other than FT that we considered for comparison: 
The no flow case (QF) and the do-nothing case (DN). 
As mentioned above, the QF case is the same as the DN case with $G \to \infty$. 
Similarly, the DN case is the FT case with $M = A = 1$ and $W/G = W^*/G^*$. 

The approximation of the energy equation, Eq.~\ref{approximate energy}, allows computer optimization of $W/G$, $M$, and $A$ to be performed using only the multivariate statistics $T(W,MSt,A)$, $\alpha_1(W,MSt,A)$, and $\alpha_2(W,MSt,A)$, and not the trajectories themselves. 
The simulations are capable only of randomly sampling from the distribution of these parameters and therefore we only estimated the underlying manifolds $T(W,MSt,A)$, $\alpha_1(W,MSt,A)$, and $\alpha_2(W,MSt,A)$. 
This results in some roughness on the discretized grid and interferes with the optimizations since it generates spurious local minima. 
Furthermore, optimization requires that the parameters are defined at all points within the grid, and not just at the grid points. 
This was accomplished by first applying a 3D Gaussian filter to each of the parameter estimates with a 1.5 grid-point standard deviation and then using a 3D spline to construct estimates of the underlying manifold and subsample from it. 
The functional in Eq.~\ref{approximate energy} was then used as the performance function for goal (1), and as the constraint for goal (2). 
Standard gradient-descent methods were sufficient because the optimization landscape was convex except at large $G$ and small $St$ for which the performance of all strategies is nearly identical anyway. 
We used MATLAB’s {\it fmincon} function for the optimizations. 
The results presented below are optimal values of $W$, $MSt$, and $A$ for each $G$ and $St$.

\subsection{Minimum energy optimization}
When range or energy efficiency are important, it is often desirable to minimize the energy, $E_{FT}$, required to travel between two points. 
Since hovering costs energy, the problem is well posed without the need to add constraints. 
The energies used in the constraint were calculated using Eq.~\ref{approximate energy}, where $T$, $\alpha_1$, and $\alpha_2$ were computed by using spline interpolation from the simulated data at the desired $W$, $MSt$, and $A$. 

\subsection{Minimum flight time optimization}
It is often important to fly between locations as fast as possible with either a maximum allowable thrust or with a given energy budget. 
We consider the second class of flight time minimization problems with the energy budget limited for instance by the size of a battery. 
Minimization was performed under the constraint $E_{FT}/E_{DN} \leq 1$, 
or $E_{FT}/E_{QF} \leq 1$, depending on which budget was of interest, $E_{DN}$ or $E_{QF}$. 
Either of these constraints was likely to be active when $T_{FT}$ was minimized, but requiring equality in the constraint would potentially result in missing solutions for which, for instance, $T_{FT}$ was minimized and $E_{FT} < E_{DN}$. 
This could occur only if at some point $\partial E_{FT}/\partial W\leq0$, 
which usually indicates that reducing thrust would increase average flight speed, 
an unlikely but theoretically feasible scenario. 

\section{Results}
\label{resultssection}
In this section we show that relative to flight through quiescent fluid, 
cyber-physical forcing (FT) generates advantages in both energy consumption and transit time, 
meaning that with appropriate forcing 
turbulence can be beneficial to flight, and not detrimental. 
To show this, 
we simulate the settling physics of particles with different properties. 
These simulations strengthen the connections between fast-tracking 
and eddy-sweeping suggested in \cite{Good} by considering a different limit, 
which is the limit of strong stiffness in the direction of flight 
rather than perpendicular to it. 
We then interpret the dynamics in the context of flight by 
computing and optimizing the power required by a flight vehicle to enact FT. 
We concentrate on the finding that relative to doing nothing (DN) to combat gusts, 
FT expands the region in parameter space within which advantages are realized. 
We discuss flight optimized to minimize energy first, 
and minimum-time optimizations after this. 



\subsection{Particle settling}
\begin{figure}
     \centering
     \begin{subfigure}[b]{0.45\textwidth}
         \centering
         \includegraphics[width=\textwidth]{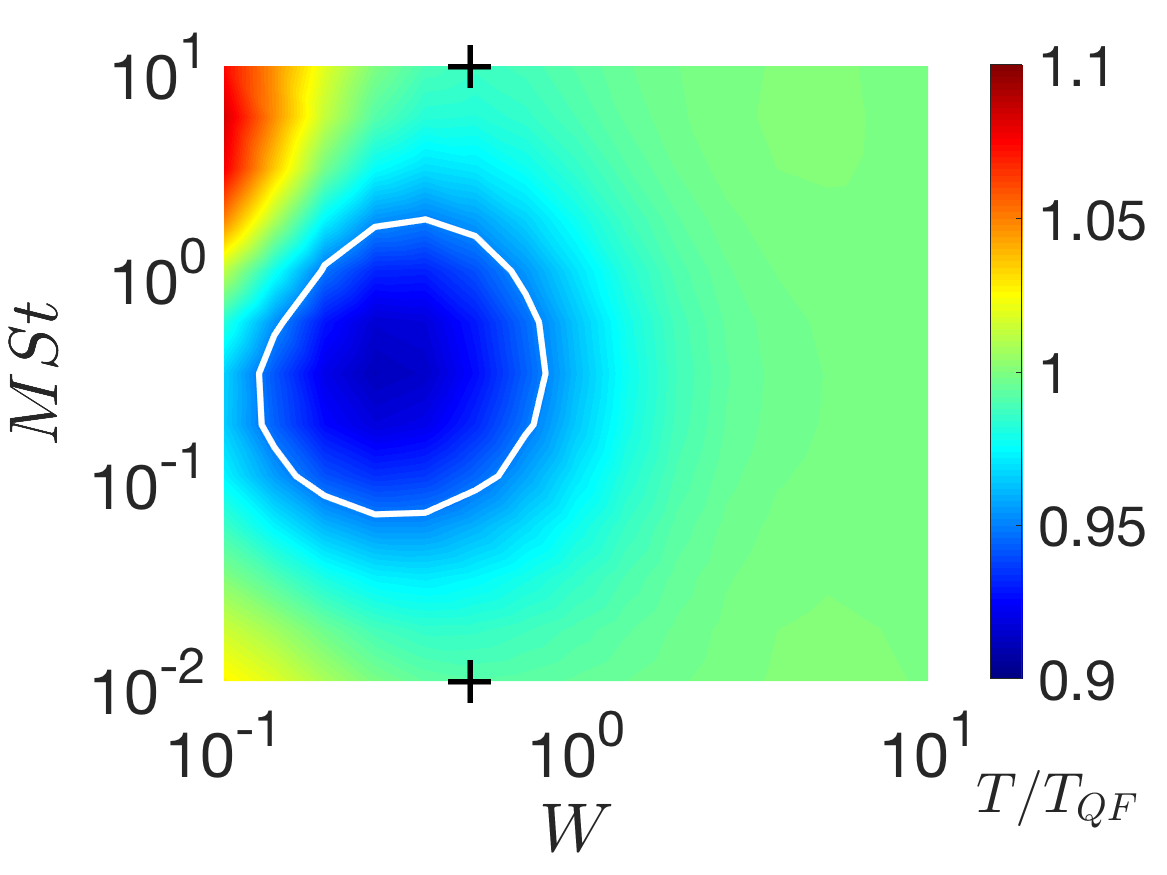}
         \caption{}
         \label{Settling time graph}
     \end{subfigure}
     \begin{subfigure}[b]{0.45\textwidth}
         \centering
         \includegraphics[width=\textwidth]{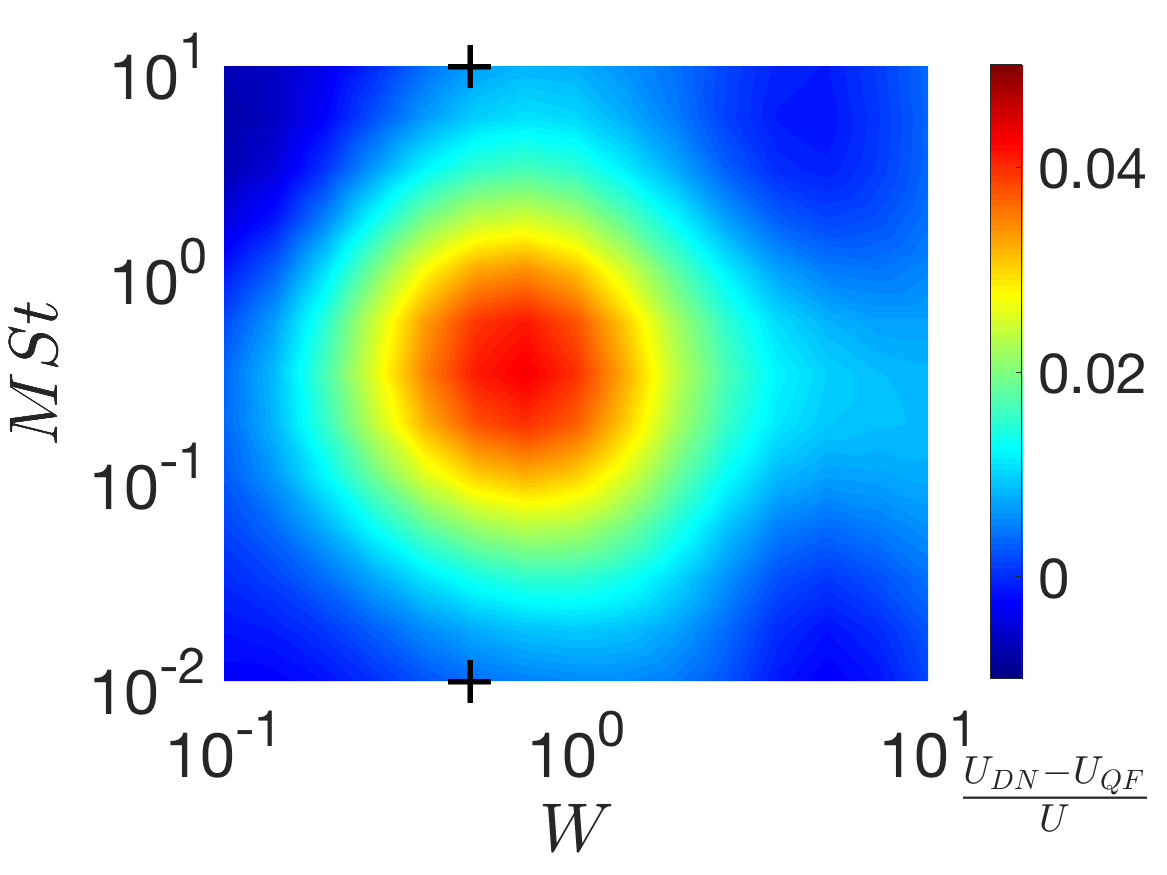}
         \caption{}
         \label{Settling enhancement graph}
     \end{subfigure}
  \caption{Either of these surfaces uniquely describe isotropic ($A=1$) 
  flight vehicle speedup caused by fast tracking. 
  For anisotropic forcing ($A \neq 1$) these are the $A=1$ slices of the corresponding manifolds. 
  The $+$'s mark the parameter values for the trajectories in Fig.~\ref{trajectories plot}. 
  (a) The particle (or flight vehicle) transit time, $T$, over a given distance, $L$, 
  through a turbulence model 
  relative to the transit time through quiescent flow, $T_{QF}$. 
  The white line delineates the region within which flight times were reduced by between 5 and 10\% relative to flight through quiescent flow. 
  (b) The corresponding mean velocity of a particle (or flight vehicle), given as the difference between the mean velocity through turbulence, $U_{DN}$, and the one through quiescent flow, $U_{QF}$, normalized by the characteristic velocity of the turbulence, $U$. 
  The speedup by turbulence vanishes in every direction away from a peak near normalized flight speeds, $W$, and normalized inertias, $M \, St$, of order one. 
  }
\label{speedup plot}
\end{figure}

We briefly review our findings that concern isotropic particles settling through turbulence, 
since the extent to which these agree with the literature benchmarks our methods. 
We use the results in Fig.~\ref{speedup plot} 
for optimization and analysis. 
Along the way we introduce particle anisotropy, 
and results that extend our understanding of settling particle physics. 

Isotropic particle settling behaviors 
agree qualitatively with experimental data, 
in the sense that there is a basin in the time required to traverse turbulent flow 
near values of the normalized flight speed ($W$) 
and normalized inertia ($St$) of order one. 
The main difference between our results and the experiments and simulations in \cite{Good} 
is that the basin we calculated is not as deep, 
as seen in Fig.~\ref{Settling time graph}. 
In our simulations, 
settling speeds are up to 5\% higher in turbulence than in a quiescent fluid 
(Fig.~\ref{Settling enhancement graph}), 
whereas \cite{Good} reports substantially larger speedups of up to 300\%, 
possibly due to weaker correlations in the model than in real turbulence. 
As discussed in Sec.~\ref{sec:flowsimulation}, 
our turbulence model is known to under-predict fast-tracking effects, 
so that this discrepancy is expected \cite[][]{WangandMaxey}. 
Our results agree quantitatively with those in \cite{Maxey87} 
for the 3D version of the same turbulence model. 

\begin{figure}
\centering

        \includegraphics[width=\textwidth]{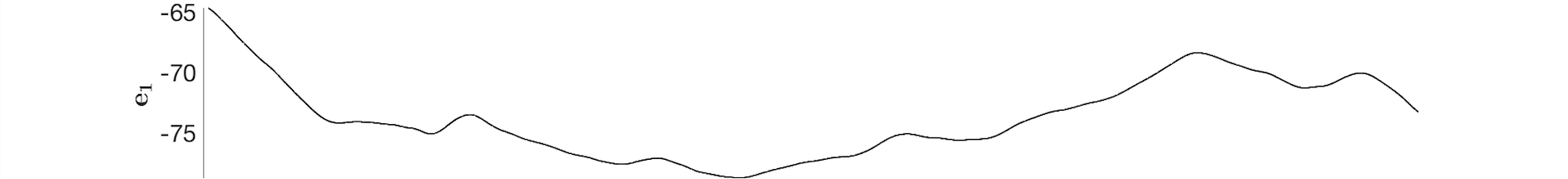}
        \includegraphics[width=\textwidth]{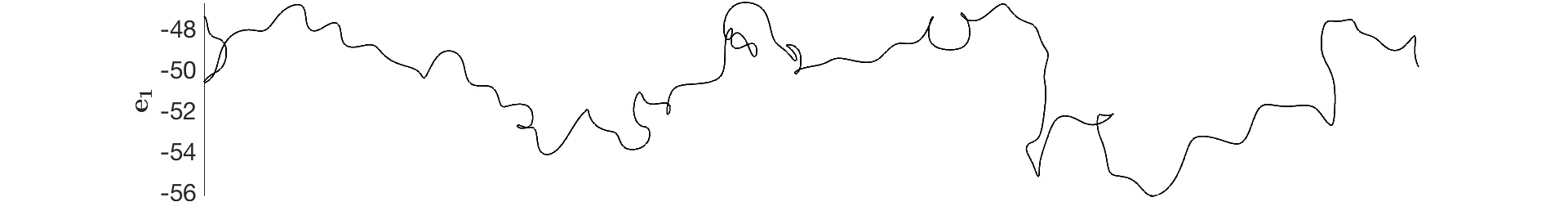}
        \includegraphics[width=\textwidth]{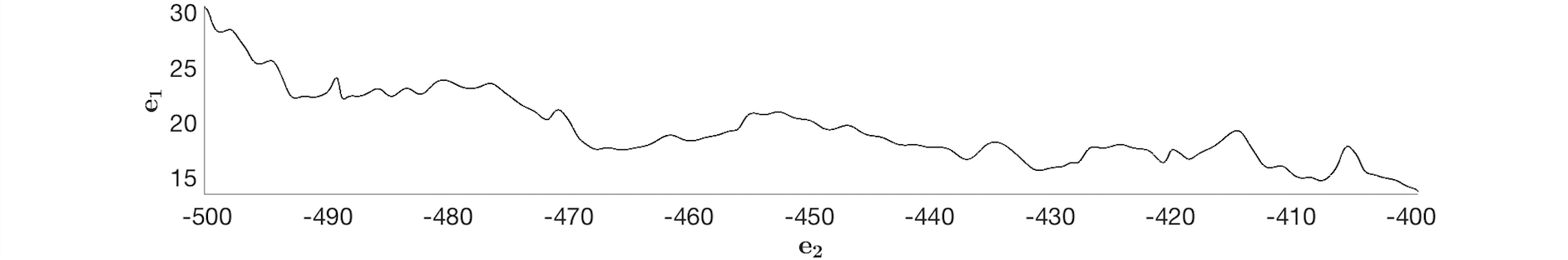}
         
    \caption{Examples of computer-simulated fast-tracking flight trajectories traveling from right to left through a turbulence model. 
    These trajectories are faster and require less energy than straight flights (not shown), which typically encounter headwinds and tailwinds with equal probability. 
    {\it Top:} Flight time is 0.3\% shorter than a straight one. 
    The inertia is isotropic and large, the normalized inertia, $St$, is 10, and the normalized flight speed of the vehicle, $W$, is 0.5. 
    {\it Middle:} $St$ is 1000 times smaller and $W$ is the same as for the above. 
    Flight time is 3\% shorter, despite the increased path length. 
    {\it Bottom:} $W$ and $St$ are the same as for the above, but an anisotropy in the virtual inertia of $A$ = 1000 reduces flight time by 13\% relative to the first example. 
}
\label{trajectories plot}
\end{figure}

Figure \ref{trajectories plot} shows that inertial anisotropy 
qualitatively changes settling behavior 
and tends to maintain 
a susceptibility to turbulent fluctuations in $\mathbf{\hat{e}}_1$ 
while reducing the overall tortuosity of trajectories. 
One way to understand this effect 
is that it may allow turbulence to sweep particles (or flight vehicles) from side to side 
into favorably moving eddies 
while simultaneously preventing the particles from back-tracking. 
As a result, 
path lengths are longer, 
though never longer than about twice the length of a straight flight, 
and the particles maintain speed toward their destinations. 
The length scales for the features in these curved trajectories 
scale with the correlation length of the turbulence. 

The limit of small anisotropy, $A \to 0$, represents trajectories confined to straight paths aligned with the direction of gravity. 
\cite[][]{Good} shows that fast tracking does not operate in this limit, 
which indicates that movement perpendicular to gravity is essential for fast tracking to work. 
The other limit, $A \to \infty$, where all acceleration is perpendicular to gravity, 
has not previously been tested. 
In our simulations of 
highly anisotropic particles ($A = 2154$), 
we find maximum settling rate enhancements 
nearly five times the isotropic value. 
Furthermore, this speedup extends to a wider range of $M\, St$, 
to the extent that we did not observe a maximum in the settling rate enhancement, 
but rather a monotonic increase as $A \to \infty$. 
Together, the results from the limits $A \to 0$ and $A \to \infty$ 
show that accelerations perpendicular to gravity 
is the dominant contributor to settling rate enhancement, 
and that accelerations in the direction of gravity 
compete with this enhancement by increasing the path lengths of particle trajectories. 

Inertial anisotropy is unphysical in the sense that particle inertia (mass) is scalar, 
but is synthesized by the cyber-physical FT. 
If anisotropy can be realized in flight vehicles in the ways discussed below, 
it could result in flight time reductions, 
which could in turn translate to energy reductions if the costs of accessing the anisotropic behavior are not too great. 
We explore these costs in the next section. 

\subsection{FT optimization}
We minimize either energy or time. 
To minimize energy, we find values of $W$, $M \, St$, 
and $A$ that minimize Eq.~\ref{approximate energy} for given values of $G$ and $St$. 
Flight trajectories that minimize energy 
are characterized by variable flight times, $T$, 
so that the energy consumption and flight time are each outputs of energy minimization. 
On the other hand, 
to minimize time we fix an energy budget, use $n = 3/4$ in Eq.~\ref{rotor power}, 
and find values of $W$, $M \, St$, and $A$ that maximize average flight speed. 
We find 
relative advantages in time that are much larger than those in energy in ways that we discuss. 

For each objective, we test two cases: one in which the inertial anisotropy was optimized, and on in which inertia was fixed and isotropic. 
The dominant feature of the optimum anisotropy that we find is an approximately diagonal line across the parameter space. 
Above the diagonal, the optimal virtual inertia is nearly isotropic, so that the flight vehicle dynamics are nearly identical to those of settling particles. 
There is a sharp transition across the diagonal to a regime where strong anisotropy in the virtual inertia became advantageous as explained below. 

The range $0.01 \leq St \leq 10$ was chosen because all significant features and changes happened in this range. 
The range $0.1 \leq G \leq 10$ was chosen because all significant features happened for $G \leq 10$, and while most significant features happened for $0.1 \leq G$, simulation time became excessive when $G$ was small.

\subsubsection{Energy minimization} 
\begin{figure}
    \centering
    \begin{subfigure}[b]{0.45\textwidth}
         \centering
         \includegraphics[width=\textwidth]{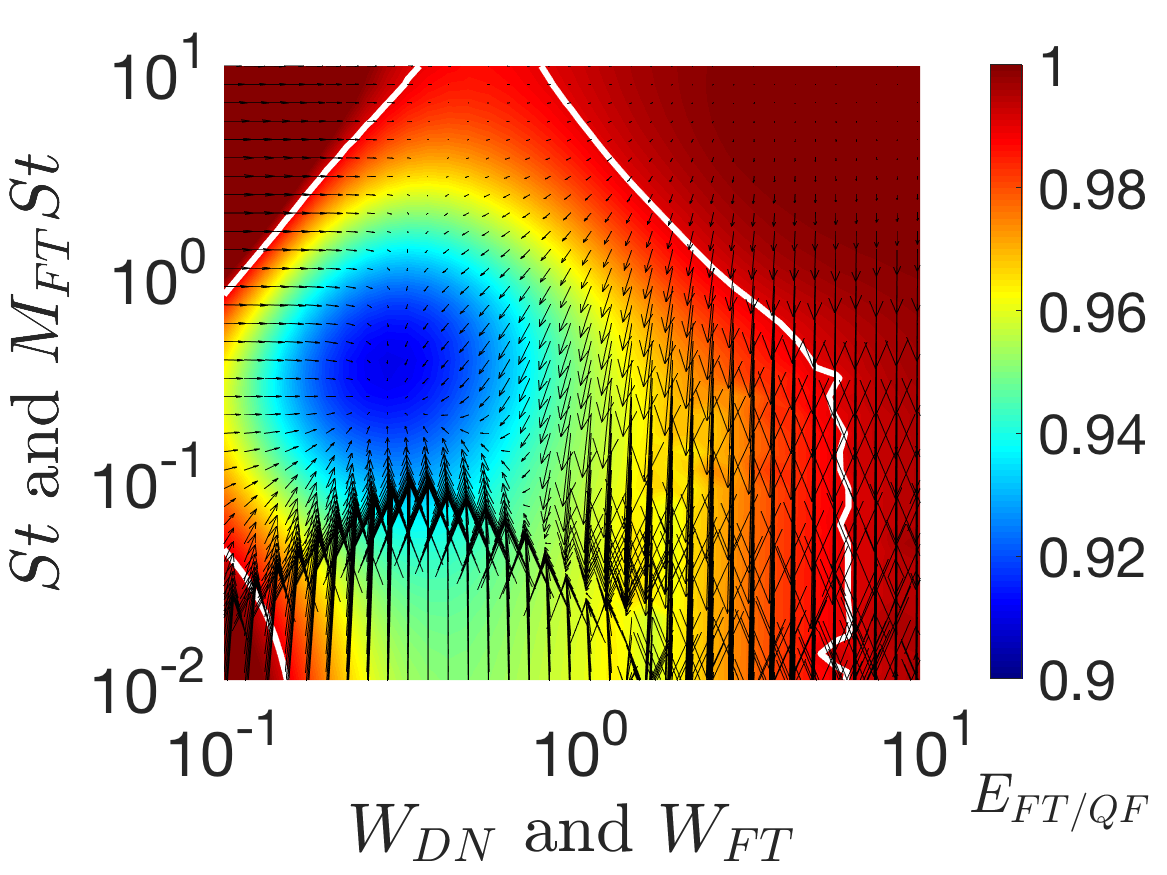}
         \caption{}
         \label{Ea}
    \end{subfigure}
    \begin{subfigure}[b]{0.45\textwidth}
         \centering
         \includegraphics[width=\textwidth]{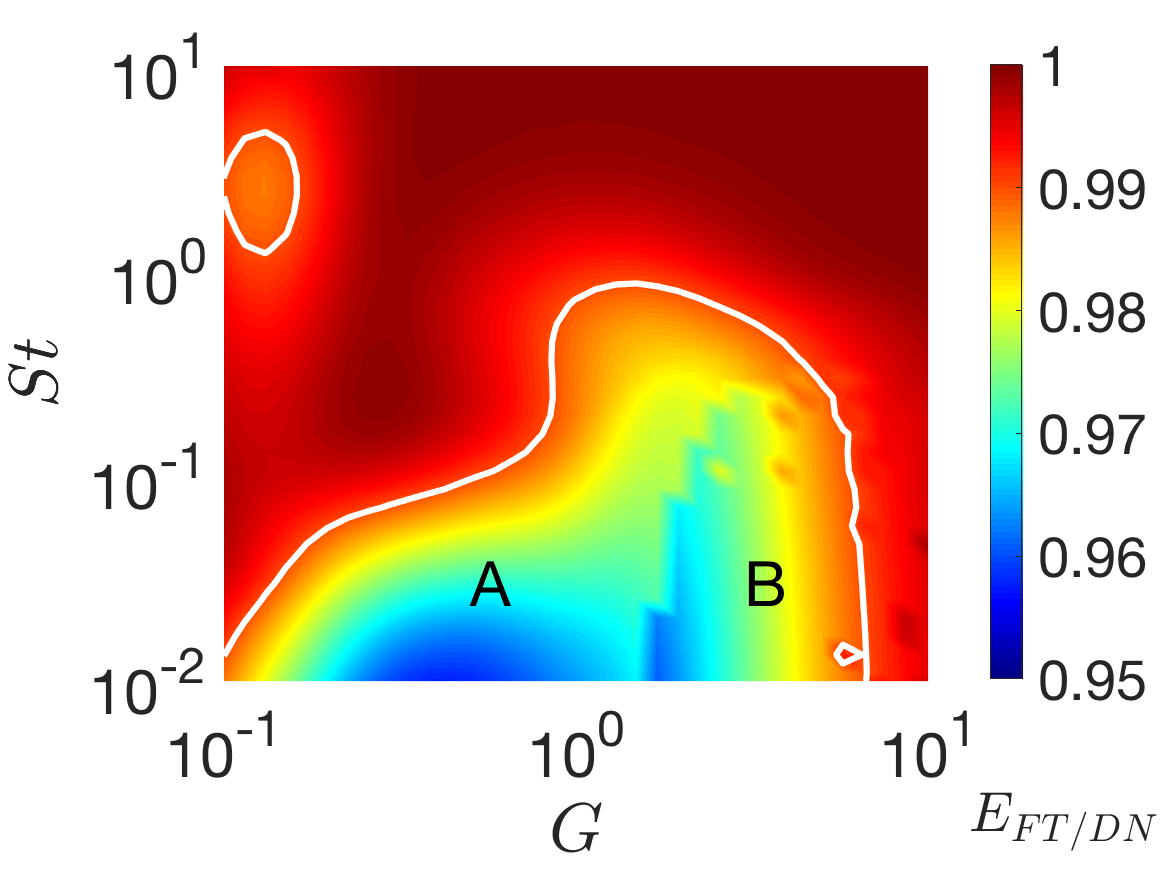}
         \caption{}
         \label{Eb}
    \end{subfigure}
    \caption{
        (a) Minimum-energy cyber-physical forcing (FT) in turbulence consumes less energy than flying through quiescent fluid (QF), and by between 1 and 10\% less within the region delineated with the white line. 
        The data are in the DN strategy space $(W_{DN},St)$ rather than in $(G,St)$ as in other plots, in order to show 
        the vector field that represents the mapping $(W_{DN},St) \to (W_{FT},M_{FT}St)$, where the tail of each vector is in $(W_{DN},St)$ (the DN strategy space), and the tip is in $(W_{FT},M_{FT}St)$ (the FT strategy space). 
        Arrows that point down correspond to FT 
        that decreases vehicle effective masses. 
        (b) FT forcing extends the advantages enabled by fast tracking beyond what 
        DN automatically realizes by between 1 and 5\% within the region delineated by the white line. 
        We expect both this region and the gains within it to be larger in real turbulence as discussed in the text. 
        Isotropic flight vehicles performed well in the region marked ``A,'' 
        while anisotropy enabled efficient flight in the region marked ``B.''
        }
        \label{Energy figures 2}
\end{figure}

Fig~\ref{Ea} displays the energy ratio $E_{FT}/E_{QF}$ for each DN strategy $(W_{DN},St)$ as a color map. 
This energy ratio represents the net energy extracted from the turbulence by the flight vehicle, in the sense that the flow energy would be smaller at the end of a flight by the relative amount $1-E_{FT}/E_{QF}$ if FT were enacted than if it were not -- note however that our model does not incorporate the effect of the flight on the flow, and the flow in our calculations was unchanged by the vehicle. 
Fig~\ref{Ea} also shows the parameter mapping, $\log{\boldsymbol{J}}$ defined by Eq.~\ref{composite mapping}, which specifies parameters for optimized FT forcing, 
and represents the forcing unique to FT. 
DN optimized as in Sec.~\ref{benchmark thrust}
results in an additional shift $W_{DN} = (2/3)\sqrt{2}G$ relative to QF. 
The isoline at $E_{FT}/E_{QF} = 0.99$ illustrates that FT is attracted to the basin of the isotropic slice of the time manifold $T$ normalized by $T_{QF}$, $(T/T_{QF})\vert_{A=1}$, centered on (0.3, 0.3) (Fig.~\ref{Settling time graph}) when the optimal $A$ is small. 
The mapping is nearly vertical for small $St$ since the cost of changing $W$ is greater than that of changing $M\, St$ there, {\it i.e.} the cost of working against drag is higher than the cost of accelerating a light-weight flight vehicle. 

Fig.~\ref{Eb} shows the extent to which FT performed better than DN shown specifically as the ratio $E_{FT}/E_{DN}$ for each set of the environmental conditions $(G,St)$. 
The benefits resulting from the attraction to the time manifold's basin 
were strongly impacted by inertia and by the behavior of the turbulence-induced accelerations (Fig.~\ref{alpha plot}), to the extent that benefits were mostly confined to $St < 1$ as shown by the isoline at $E_{FT}/E_{DN}=0.99$. 
When restricted to isotropic inertia, the benefits were further restricted to $G<1$, a region denoted ``A'' in Fig.~\ref{Eb}. 
Anisotropy did not appreciably change the maximum performance of FT. 
However, the introduction of anisotropy extended advantages to both larger $St$ and into a region of larger $G$, marked ``B'' in the figure, that includes values of $G$ approximately three times larger than for isotropic forcing. 
It is crucial for any strategy to perform well at higher $G$ because this is where the majority of atmospheric applications lie. 

Anisotropic inertia may permit fast-moving flight vehicles to move side-to-side and so to hop from one vortex to the next without expending energy and time moving fore-and-aft. 
For slower flight vehicles, trajectories are increasingly tortuous, and accelerations in both directions may be needed to find paths through favorable winds. 
The forces required to produce anisotropy become prohibitively costly to produce at low flight speeds. 
We expect realistic settling-time basins to extend the advantages to yet larger $St$ and $G$ and to a certain extent to close the annulus around the basin. 

\begin{figure}
     \centering
     \begin{subfigure}[b]{0.32\textwidth}
         \centering
         \includegraphics[width=\textwidth]{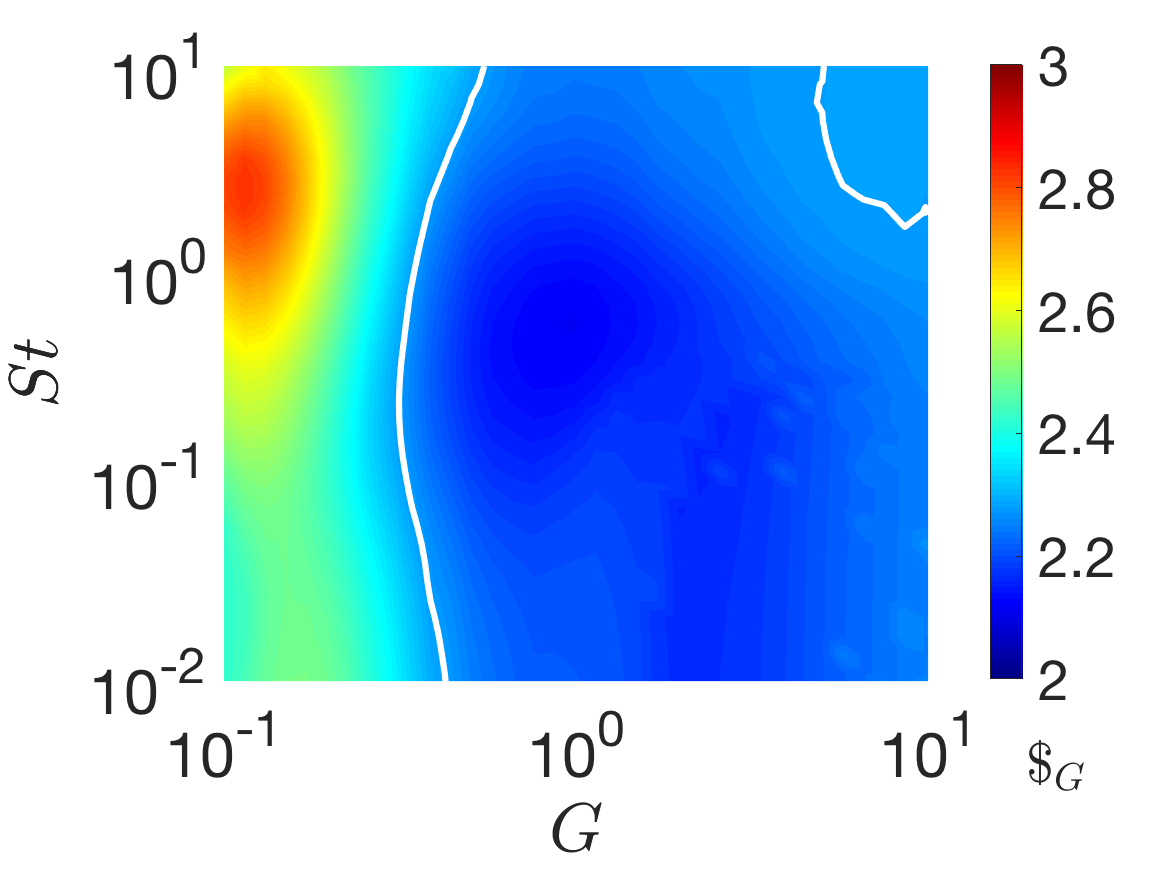}
         \caption{}
         \label{Ee}
     \end{subfigure}
     \begin{subfigure}[b]{0.32\textwidth}
         \centering
         \includegraphics[width=\textwidth]{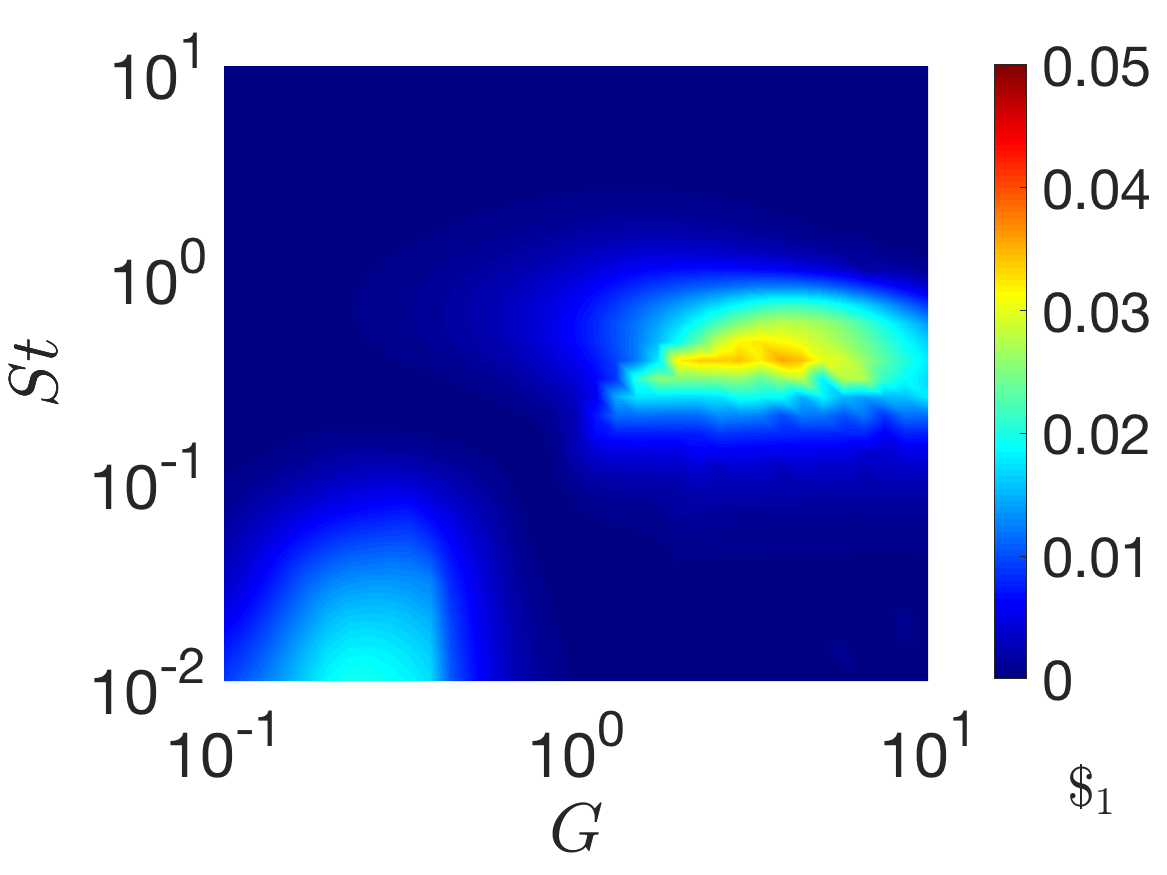}
         \caption{}
         \label{Ef}
     \end{subfigure}
     \begin{subfigure}[b]{0.32\textwidth}
         \centering
         \includegraphics[width=\textwidth]{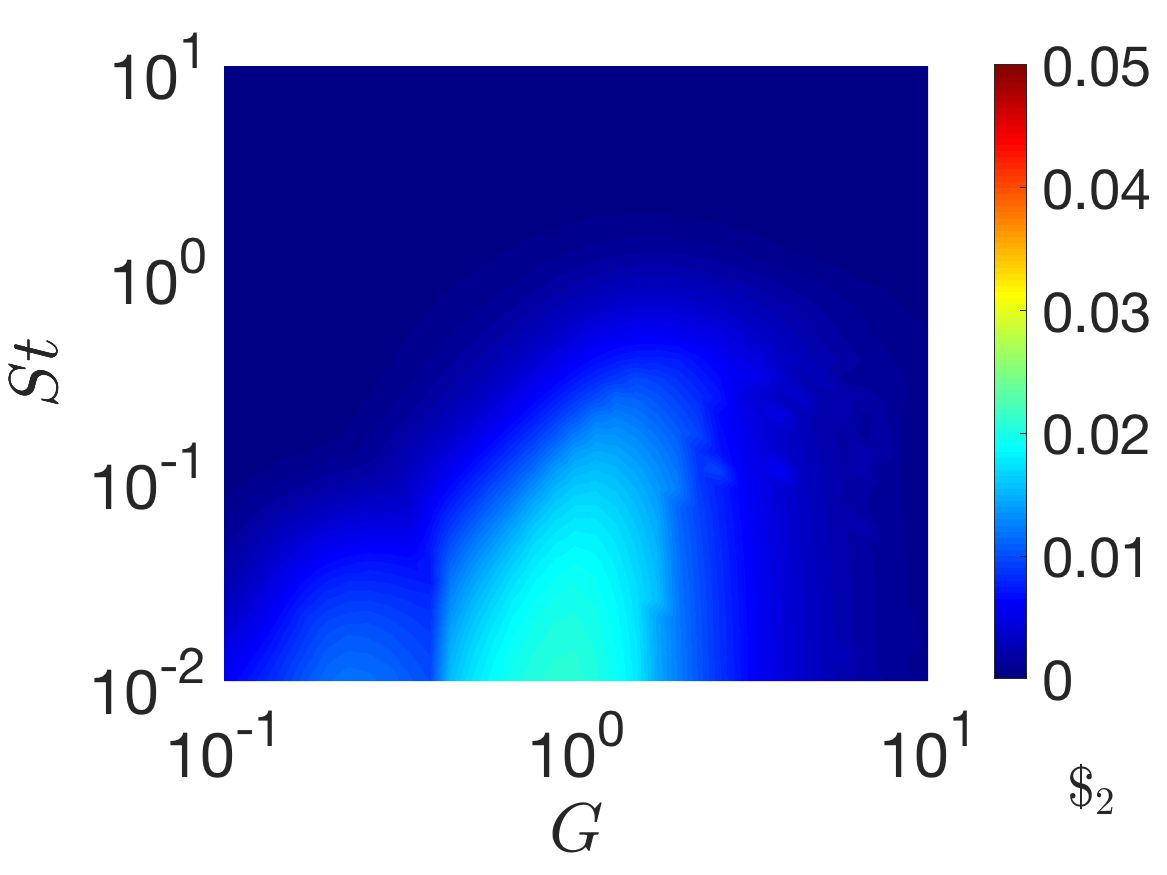}
         \caption{}
         \label{Eg}
     \end{subfigure}
    \caption{
        Costs associated with minimum-energy FT. 
        (a) The dimensionless energy spent to keep the vehicle aloft 
        (working against $\boldsymbol{g}$) 
        and to push it toward its destination (with thrust ${f}_0$) is larger for slow-moving flight vehicles in part because they spend more time aloft. 
        The white line separates larger (red) and smaller (blue) thrust than is optimal in quiescent fluid. 
        (b) Dimensionless power expended to accelerate
        transverse to the mean flight direction, 
        which tends to shift vehicles between vortices. 
        (c) Dimensionless power expended to accelerate 
        toward the destination. 
    }
    \label{min E cost figures}
\end{figure}

The energetic costs of FT forcing, $\$_G$, $\$_1$, and $\$_2$ 
in Eq.~\ref{Flight Power Components}, 
can be seen in Fig.~\ref{min E cost figures}. 
For these solutions the anisotropy was free to take whichever value resulted in the lowest energy. 
Fig.~\ref{Ee} shows that the combined cost of staying aloft and producing the destination-seeking thrust, $\tilde{f}_0$, was often significantly different from its dimensionless DN value of $\$_G=3^{3/4}$ (included as an isoline). 
In other words, turbulence leads to a different optimal thrust in a nontrivial way. 
The cost of modifying $M$ to produce desired accelerations in $\mathbf{\hat{e}}_1$, $\$_1$, and in $\mathbf{\hat{e}}_2$, $\$_2$, are both generally small and restricted to small $G$ and small $St$ (Figs.~\ref{Ef} and \ref{Eg}, respectively). 
The fact that these costs are small justifies the use of the energy approximation, Eq.~\ref{approximate energy}, as discussed in the theory section. 
The costs $\$_1$ and $\$_2$ for a given $M$ and $A$ both scale with $St^2/G^2$, while $\$_G>1$ regardless of $G$, so that modifying the dimensionless inertia components through $MA$ and $M$ become prohibitively costly for large $St/G$. For small $St/G$, the costs of changing $M$ and $MA$, $\$_1$ and $\$_2$, are smaller so that more energy is allocated to the forcing 
than to the thrust. 
As a result, changes to $M$ and $MA$ dominate the behavior there as discussed below. 

\begin{figure}
     \centering
     \begin{subfigure}[b]{0.32\textwidth}
         \centering
         \includegraphics[width=\textwidth]{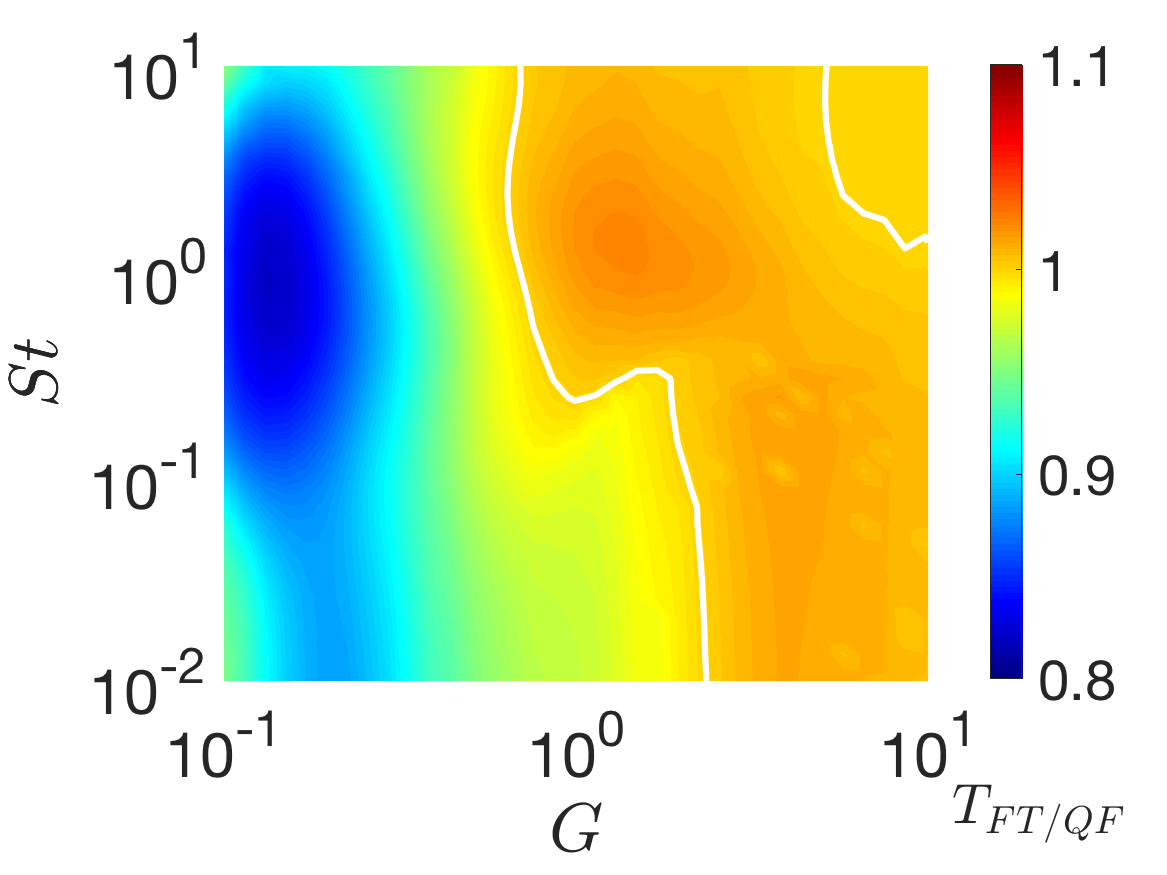}
         \caption{}
         \label{Ec}
     \end{subfigure}
     \begin{subfigure}[b]{0.32\textwidth}
         \centering
         \includegraphics[width=\textwidth]{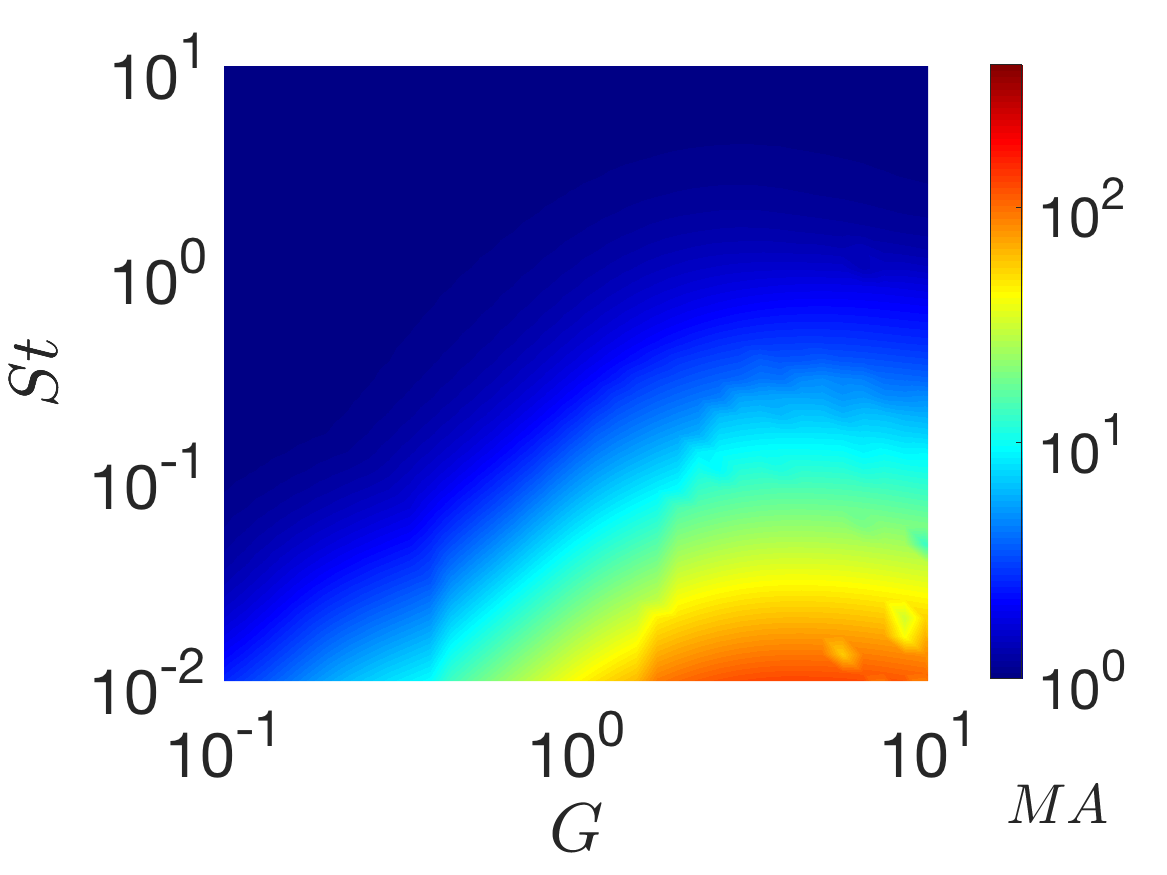}
         \caption{}
         \label{Eh}
     \end{subfigure}
     \begin{subfigure}[b]{0.32\textwidth}
         \centering
         \includegraphics[width=\textwidth]{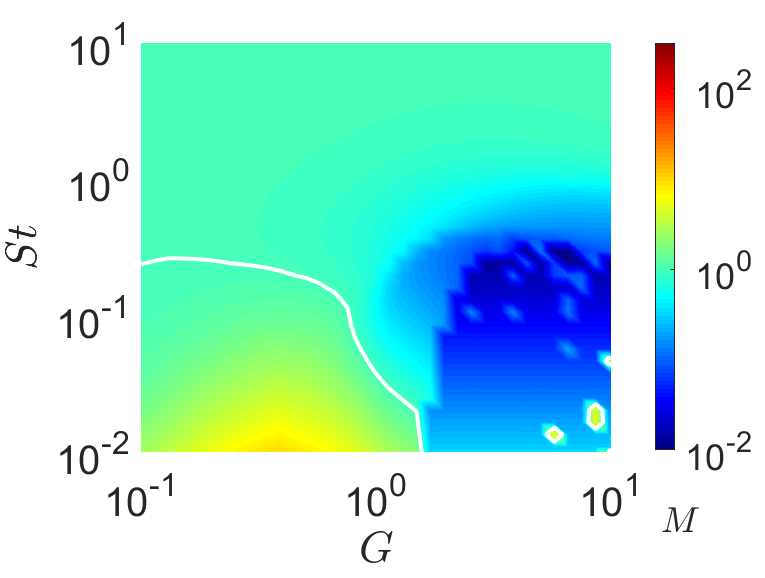}
         \caption{}
         \label{Ed}
        \label{Energy figures 1b}
     \end{subfigure}
    \caption{
    (a) To minimize energy under cyber-physical forcing (FT), it can be beneficial to take a longer time to fly through turbulence, $T_{FT}$, than through quiescent fluid, $T_{QF}$. 
    The white line separates extended flights (red) from shorter ones (blue).
    (b) In contrast, it is always favorable under FT forcing to reject disturbances in the direction of flight by increasing effective inertia in that direction ($MA$), especially for fast, lightweight vehicles. 
    (c) The effective mass in the direction transverse to mean flight ($M$), 
    is reduced under FT forcing for all but slow, lightweight vehicles. 
    }
    \label{min E time and inertia figures}
\end{figure}

A main feature of the flight time shown in Fig.~\ref{Ec} is a reduction in mean speed for $G$ greater than the value that minimizes $T_{FT}/T_{QF}$. 
Restricting the forcing to be isotropic did not affect speedup but did increase the extent of the region of reduced speed. 
The source of these changes in speed come from $W_{FT} \neq (2/3)\sqrt{2}G$, 
which corresponds to FT allowing the flow to advance the vehicle toward the destination in exchange for a change in flight speed. 
In the upper-left quadrant of Fig.~\ref{Eb} where energetic benefits are relatively small the time of flight seen in Fig.~\ref{Ec} is nonetheless substantially reduced, meaning that even when energy cannot be much reduced, a flight vehicle can nonetheless reach its destination more quickly. This speedup occurs for $G$ less than the value that minimizes $T_{FT}/T_{QF}$. We explore this finding further in the next section. 

The combination of the vector field in Fig.~\ref{Ea} and the surface in Fig.~\ref{Eh} forms a complete set of instructions for an FT controller. 
Note that the information in Fig.~\ref{Ed} is already contained in the vector field in Fig.~\ref{Ea}, and we include it here simply for comparison. 
For those vehicles in the lower right quadrant 
the inertial anisotropy, $A$, which is the ratio of the surfaces in Figs.~\ref{Ed} and \ref{Eh}, was large. 
For those vehicles with reduced effective mass, blue in Fig.~\ref{Ed}, energetically favorable trajectories are more tortuous than they otherwise would have been because FT amplified disturbances in order to hop from one vortex to another. 

\subsubsection{Time Minimization}
The energy budget was chosen to be the constant $E_{DN}$ in this section. 
The choice of energy budget did not alter the results qualitatively, but is nonetheless worth consideration. 
An energy budget of $E_{DN}$ generally reveals the particular effectiveness of FT forcing relative to DN, which appears as regions where $T_{FT} < T_{DN}$. 
As a result, time improvements can be realized with this energy budget even when drag non-linearity causes particle settling rate reductions rather than enhancement. 
An energy budget of $E_{QF}$ instead reveals the extent to which FT allows the vehicle to benefit from gusts, and is relevant when the energy budget represents a vehicle’s battery capacity, for instance, or where the goal is to determine how much more quickly the vehicle can make a route as a result of the gusts. 

\begin{figure}
     \centering
     \begin{subfigure}[b]{0.45\textwidth}
         \centering
         \includegraphics[width=\textwidth]{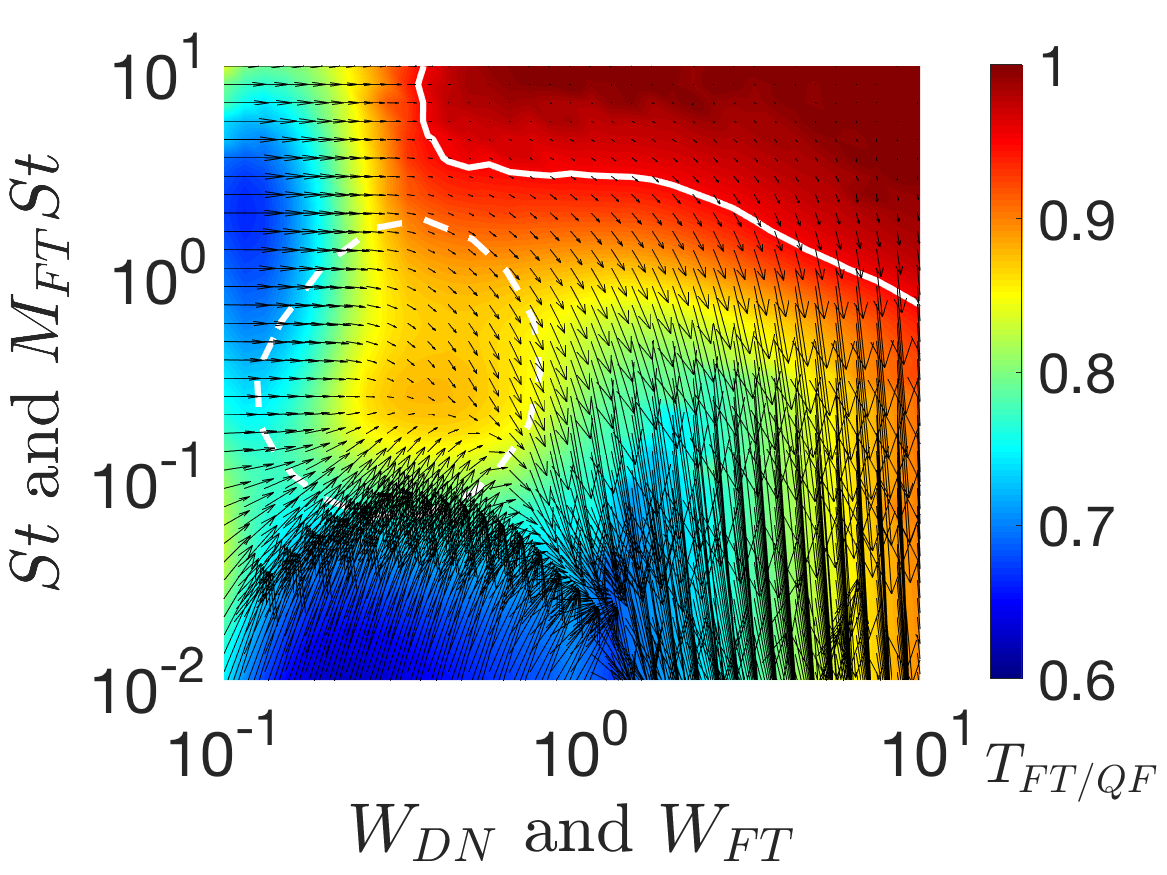}
         \caption{}
         \label{T_FT/QF}
     \end{subfigure}
     \begin{subfigure}[b]{0.45\textwidth}
         \centering
         \includegraphics[width=\textwidth]{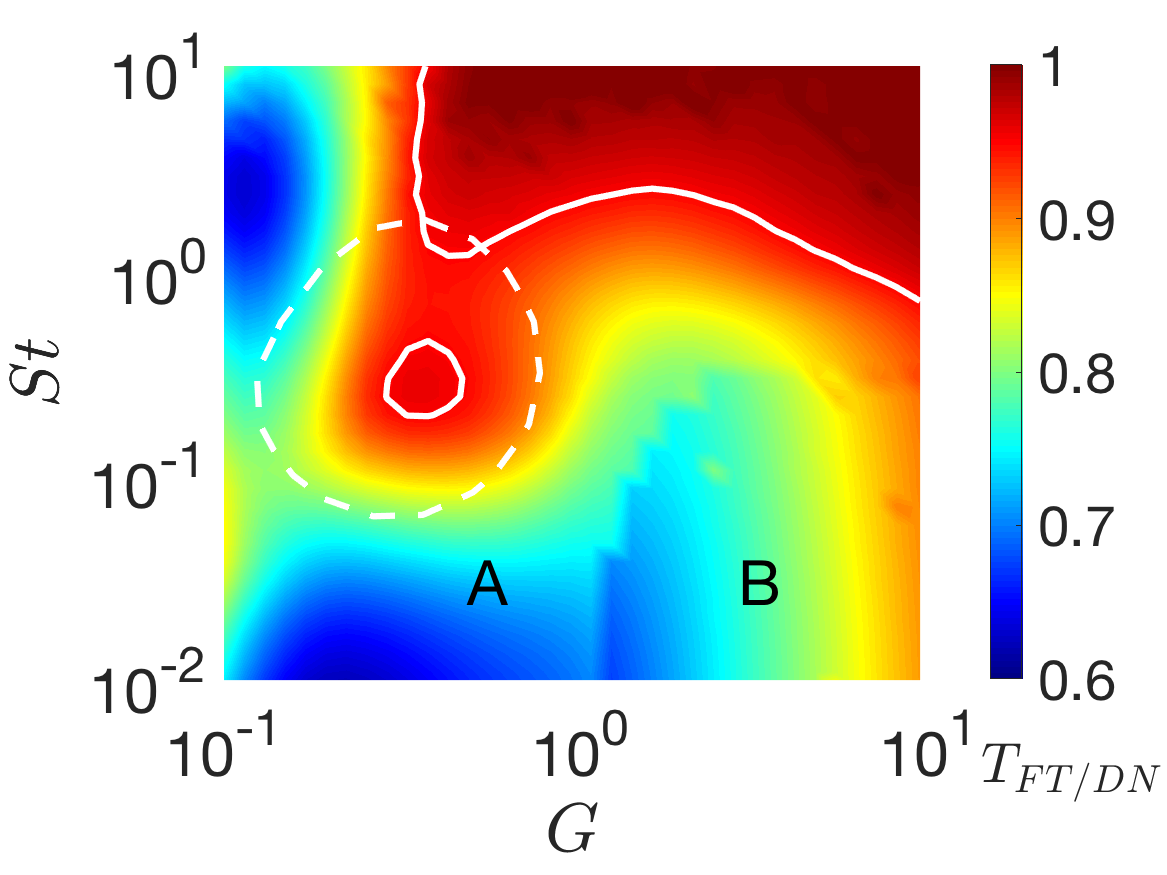}
         \caption{}
         \label{T_FT/DN}
     \end{subfigure}
    \caption{Same as Fig.~\ref{Energy figures 2}, but for minimum-time forcing (FT). 
    (a) FT in turbulence reduces flight times relative to flight through quiescent fluid (QF) by up to 40\% 
    with energy budgets given by $E_{DN}$. 
    (b) FT expands the basin of reduced flight time beyond what DN realizes automatically by fast tracking (see Fig.~\ref{speedup plot}b). 
    ``A'' and ``B'' are as in Fig.~\ref{Energy figures 2}. 
    The isolines at $0.95$ delineate regions where FT reduces relative flight times by more than 5\%. 
    The dotted lines are the corresponding isolines without FT from Fig.~\ref{speedup plot}b for comparison. 
    }
    \label{time mapping}
\end{figure}

FT achieved much greater advantages in flight time than in energy in a given flow. 
As seen in Fig.~\ref{T_FT/QF}, the time ratio $T_{FT}/T_{QF}$ was as low as 0.6, whereas the corresponding energy ratios were only as small as 0.95. 
As discussed below, this is partly a consequence of the quadratic dependence of the energy, $E$, on flight speed, $W$, near $W^*/G^*$ independent of $n$ (as can be seen in Eq.~\ref{benchmark energy}). 
As a result, small changes in $E$ can be converted into larger changes in flight speed, $W$, which in turn leads to shorter flight times. 
In Fig.~\ref{T_FT/QF}, the net rightward mapping compared with Fig.~\ref{Ea} is due to this conversion from energy saved and increased flight speed. 
The isoline at $T_{FT}/T_{QF} = 0.95$ extends beyond $W_{DN} = 10$, indicating higher performance even at flight speeds large relative to the speed of the turbulent fluctuations. 
Since the time advantages are so large, the comparison with QF and DN (Fig.~\ref{T_FT/DN}) flight times are similar. 
It is also the case that the $MA$ and $M$ surfaces under time minimization are similar to those for energy minimization (Figs.~\ref{Eh} and \ref{Energy figures 1b}, respectively). 

\begin{figure}
     \centering
          \begin{subfigure}[b]{0.32\textwidth}
         \centering
         \includegraphics[width=\textwidth]{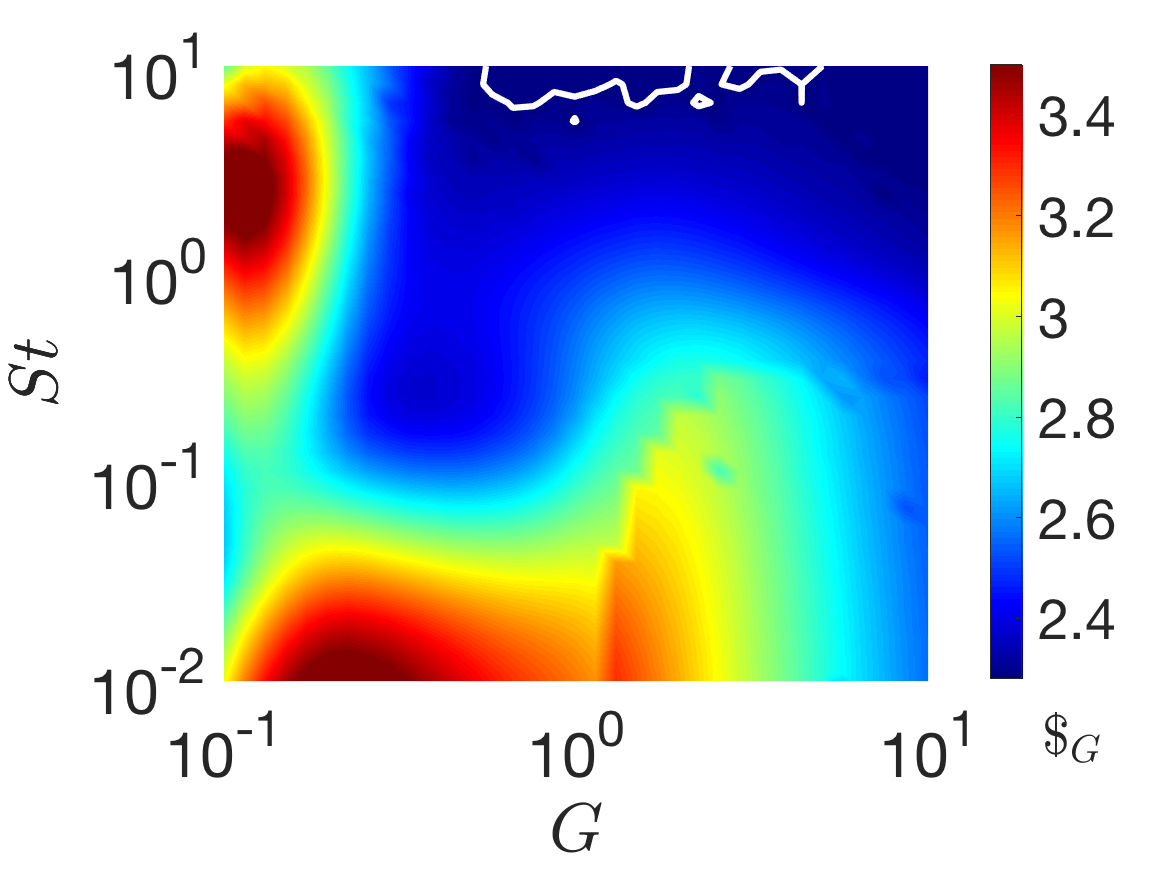}
         \caption{}
         \label{Te}
     \end{subfigure}
     \begin{subfigure}[b]{0.32\textwidth}
         \centering
         \includegraphics[width=\textwidth]{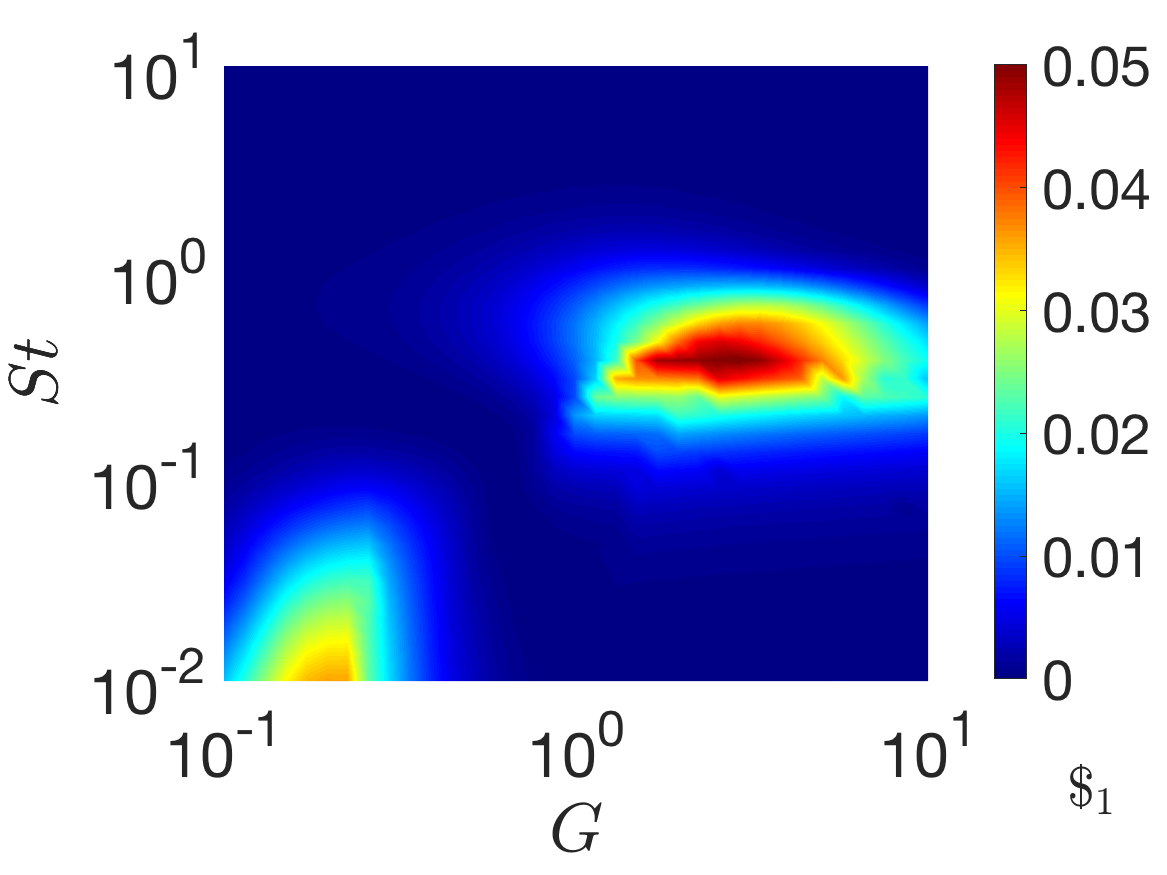}
         \caption{}
         \label{Tf}
     \end{subfigure}
     \begin{subfigure}[b]{0.32\textwidth}
         \centering
         \includegraphics[width=\textwidth]{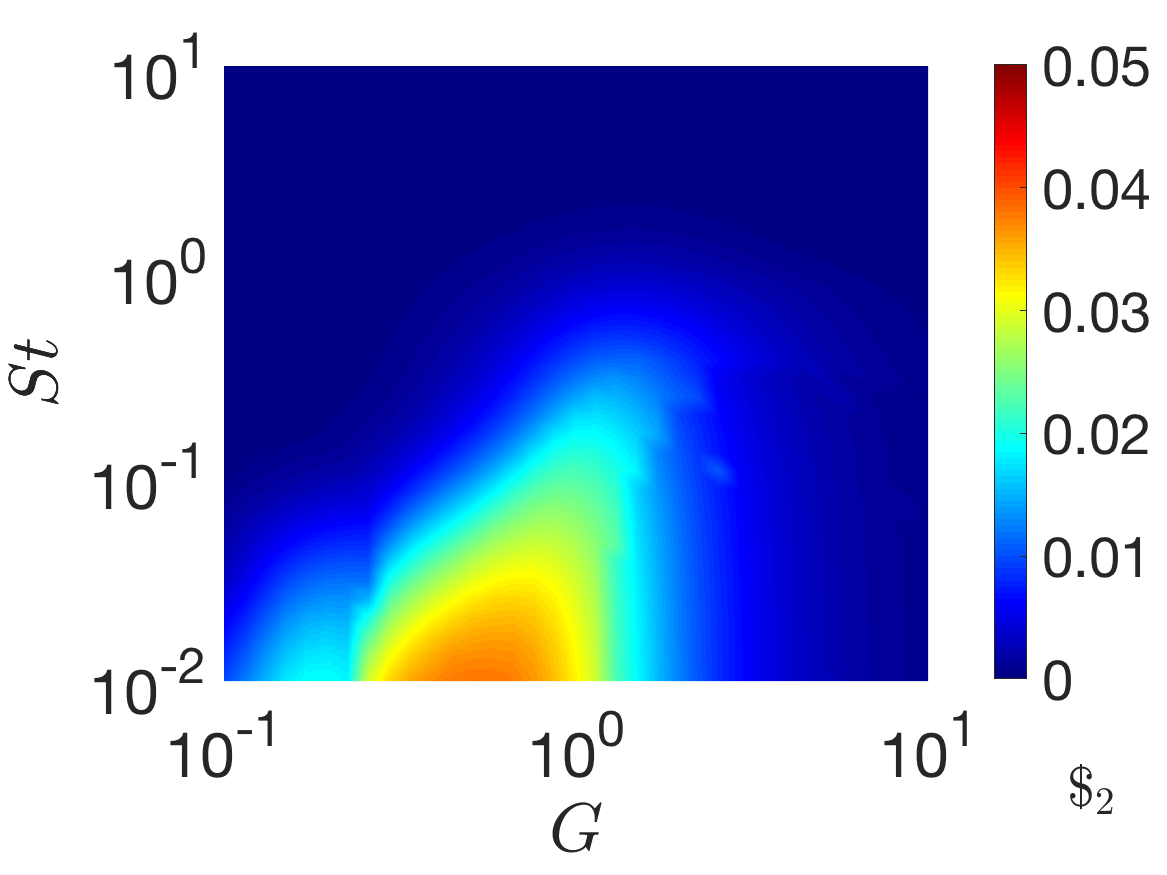}
         \caption{}
         \label{Tg}
     \end{subfigure}
        \caption{
        Same as Fig.~\ref{min E cost figures}, but for minimum-time FT using a fixed energy budget, $E_{DN}$. 
        (a) The combined dimensionless energy spent to keep the vehicle aloft and push it toward its destination 
        is qualitatively different than for mimimum energy flight (Fig.~\ref{Ee}) 
        as discussed in the text. 
        The white line separates larger (red) and smaller (dark blue) thrust than is optimal for QF. 
        The dimensionless power expended on accelerations 
        transverse to the mean flight direction (b), 
        and in the mean flight direction (c), 
        were both similar to the minimum energy solutions. 
        }
        \label{Time figures}
\end{figure}

The energetic costs under anisotropic FT forcing are qualitative similar for 
time (Fig.~\ref{Time figures}) and energy (Fig.~\ref{min E cost figures}) minimization, 
except that the forcing is more active while minimizing time since there is more energy available to the forcing 
-- the energy budget is expended entirely during flight and there is no advantage to reducing energy consumption. 
This can be seen as longer vectors in Fig.~\ref{T_FT/QF} and stronger peaks in Figs.~\ref{Te}, \ref{Tf} and \ref{Tg}. 
Furthermore, since the budget is $E_{DN}$, FT does not perform much better than does DN at the point where DN performs at its best, that is, near the basin in the time manifold. 

Anisotropic FT improves performance 
relative to isotropic FT 
in a way that is practically important, 
since atmospheric applications often lie 
in the regime of large $W_{DN}$ (and $G$) where variable anisotropy is most beneficial. 
Anisotropic FT extends not only extends the region of significant benefits 
in $W_{DN}$ (by factor of about three ), 
but also toward larger $St$. 
As noted above, real turbulence likely further expands the region 
compared with the one produced by the turbulence model we studied. 

\subsection{Comparison between energy and time minimization}
Time minimization at constant energy generally resulted in greater benefits than energy minimization since the landscape of energetic costs is relatively flat with respect to changes in flight speed. 
This can be seen from a generalization of drag, Eq.~\ref{quaddrag}, to a nonlinear one given by $\tilde{\boldsymbol{f}}_{d} = k(\tilde{\boldsymbol{w}}-\tilde{\boldsymbol{u}})||\tilde{\boldsymbol{w}}-\tilde{\boldsymbol{u}}||^{p-1}/\tau_d$ for drag constant $k$, and $W/G$ redefined as $W^p/G^p=\tilde{f}_0/g$, so that Eq.~\ref{benchmark energy} now reads 
\begin{equation}
    E_{QF} \sim (G/W)\left(1 + W^{2p}/G^{2p}\right)^n. 
\end{equation}
The problem we considered heretofore was a linear approximation to the quadratic case for which $p=2$. 
For any $n$, the minimum energy occurs at $W^*/G^* = \sqrt[2p]{1/(2np-1)}$. At this point the slope $dE_{QF}/d(W/G)$ is zero so energy is quadratic with $W/G$ to leading order for any $n$. 
For nonlinear drag ($p=2$) and $n=3/4$, for example, flying 10\% faster requires only 1.3\% more energy. 
For linear drag ($p=1$), 
a 5\% energy reduction can be converted into an approximately 50\% flight time reduction by applying the energy to thrust. 
It holds generally that small energetic benefits can be converted into significant time savings. 

\section{Discussion}
\label{discussionsection}


We discuss extensions to FT that incorporate time-dependence 
and correlations between responses in different directions. 
We note that it is possible to employ FT on vehicles besides rotorcraft, 
and we explore an analogy between anisotropy in inertia and in aerodynamic drag that may facilitate the use of FT on fixed-wing aircraft and neutrally-buoyant vehicles such as submarines or blimps, 
and contribute to a better understanding of the gliding behavior of volant lifeforms. 

 

We briefly address here the implementation of FT, 
which in practice requires not only an understanding of sensor noise and various forms of delay that may limit the realizable values of $\mathsfbi{C}$ for which the vehicle is stable \cite[][]{Williamsoncyber}, 
but also an understanding of control methods 
such as implicit model following that may in turn 
expand the range of $\mathsfbi{C}$ within which stability prevails \cite[][]{Hengyemodelfollowing}. 
Furthermore, the parameters $G$ and $St$ must be known in order to implement FT. 
While the characteristics of the flight vehicle, embodied in $\tau_d$, may be known {\it a priori}, 
both the characteristic flow time scale $L/U$ and speed $U$ need to be estimated \cite[][]{Morelli2012,Al-Ghussain2020,Gonz_lez_Rocha_2020,HengyeparameterID}. 
This may be accomplished using local weather information or on-board accelerometers and by measuring the characteristics of correlation functions. 
When $M\,St > 1$ the flight vehicle's inertia attenuates its response to flow structures \cite[][]{Warhaftparticleacceleration}, 
so that an understanding of the vehicle's dynamics is needed in practice to infer $G$ and $St$ from accelerometry alone. 

Concerning anisotropy, FT always finds nearly isotropic inertia, $A \approx 1$, 
to be optimal unless $G>1$ and $St<1$. 
For $G>1$, the behavior changes abruptly because it becomes energetically favorable to mimic anisotropic behavior. 
To see why the change is abrupt, 
consider that when $A = 1$, the optimal $M$ and $W$ are values for which the flight time is reduced and the forcing is small. 
This is achieved by traveling in parameter space towards the time minimum, (seen in Fig.~\ref{Settling time graph}). 
When $St$ is small enough that the vehicle is below the basin in parameter space, 
minimizing flight time means making $M>1$ in order to push the virtual inertia upward toward the basin, which corresponds to a nearly vertical vector field, as seen in Fig.~\ref{Ea}. 
However, once anisotropic strategies become energetically favorable, the better strategy is to make $M<1$ so as to reduce $MA$. 
This permits $A$ to be made larger, allowing access to more favorable slices of the time manifold without incurring an excessive increase in the cost component $\$_2$ (since $\$_2\propto (1-MA)^2$). 
Because the isotropic locally optimal behavior requires $M>1$ while the anisotropic locally optimal behavior requires $M<1$, and $M\sim 1$ is not optimal, 
the transition between these two local optima is necessarily abrupt.

\subsection{Extensions and modifications to FT forcing}
Changes to fast-tracking forcing (FT), beyond the introduction of inertial anisotropy, 
have the potential to further improve performance by exploiting anticipated regularities in the structures of the flows a vehicle traverses. 
Some of these modifications can be implemented without the need for flow measurements, 
as is the case for the FT forcing analyzed in this paper. 
There is useful information contained in the history of a particle's trajectory, including its angular accelerations \cite[e.g.][]{VothAnisotropicParticle}, which is ignored by our model but offers potential for further development. 
In the example below, we allowed the effective mass to be variable in flight, 
and to be coupled in different directions.

To avoid loitering near stagnation points, inertia transverse to the direction motion can be assigned a linear function of the acceleration in the direction of motion. 
Deceleration in the direction of mean motion would then tend to push the vehicle off-track temporarily, by reducing the effective mass, in order to avoid the potential loiter-inducing structure. 
Conversely, in vortices that push the vehicle toward the destination, the vehicle would accelerate in the direction of motion. 
As a result, the particle inertia transverse to the direction of motion would increase, possibly causing the vehicle to loiter beneficially in these parts of the flow. 
The intended effect of the coupling is to cause vehicles to seek out areas of high tailwind velocity and avoid stagnation points, in contrast to particle-like behavior that tends to concentrate particles and vehicles in areas of high strain rate and low vorticity.

\subsection{Generalization and application of FT to other vehicles}

FT forcing for rotorcraft maneuvering in two dimensions (2D), which is what we analyzed, 
generalizes to three dimensions and to other vehicles. 
These include fixed-wing aircraft and neutrally-buoyant vehicles like blimps, unmanned submersibles, and ships (ships being neutrally buoyant but constrained to 2D). 
The point-mass assumption is useful and usually applicable \cite[e.g.][]{Patel2006sine,crazyflypointmass}, 
however the forces are different from those considered in this paper, and often more complex. 
For example, fixed-wing aircraft experience a lift and drag force dependent on their airspeed and angle of attack. 
This introduces additional non-linearities in the dynamics. 



Of note is the fact that aerodynamic anisotropy associated with an asymmetry in the vehicle geometry or aerodynamics appears in our model in the same way as inertial anisotropy. 
That is, fixed-wing aircraft, for which the lift-to-drag ratio is usually high, 
behave anisotropically, in the sense that $A>1$, without the need for forced adjustment of $A$. 
As a result of their natural anisotropies, FT may be more effective when employed on these vehicles, and we formalize this in what follows. 

For birds or fixed-wing aircraft in level flight at speed $U_{QF}$ with constant lift and drag coefficients and with lift-to-drag ratio ${L/D}$, a small horizontal gust of speed $\tilde{\delta}_2$ causes accelerations of the form  
\begin{equation}
    \frac{1}{g}\frac{d\tilde{\boldsymbol{u}}}{d\tilde{t}} = 2\frac{\tilde{\delta}_2}{U_{QF}}\begin{bmatrix}
    1\\
    (L/D)^{-1}
    \end{bmatrix}.
\end{equation}
The appearance of $g$ comes from the fact that level flight requires a bird or aircraft to generate enough lift to support its weight. 
If instead the small gust were vertical, and the coefficient of lift is determined by a linear function of angle of attack, then the lift slope, $S_l$, which is $2\upi$ for an infinite wing, and the lift-induced drag would change with the gust. 
For spanwise efficiency $e$ and aspect ratio $AR$, the resulting acceleration is approximately
\begin{equation}
        \frac{1}{g}\frac{d\tilde{\boldsymbol{u}}}{d\tilde{t}} = \frac{\tilde{\delta}_1}{U_{QF}}\begin{bmatrix}
    {S_l}/{C_L} + (L/D)^{-1}\\
    {2S_l}/({\upi e AR}) - 1
    \end{bmatrix}. 
\end{equation}
The response is a sum of two terms for vertical gusts because the lift and drag vectors are both rotated by the change in apparent wind angle. 
Together, these gust-acceleration relations mean that for small gusts perturbing steady flight at $U_{QF}$, the dimensionless equation of motion for a bird or fixed-wing aircraft is, 
\begin{equation}
\frac{d\boldsymbol{u}}{dt} =\frac{G}{W_{QF}St} \begin{bmatrix}
    {S_l}/{C_L} + (L/D)^{-1}  &  2\\
    {2S_l}/({\upi e AR}) - 1   &  2{({L/D})^{-1}}
    \end{bmatrix}(\boldsymbol{w}-\boldsymbol{u}+\hat{\boldsymbol{e}}_2).
    \label{fixed-wing dynamics}
\end{equation}
This equation of motion
is similar to the one for a point particle (Eq.~\ref{FT Dynamics}), except that $U/g$ rather than $\tau_d$ characterizes gust response time, and the off-diagonal terms are non-zero and not equal -- the matrix is not symmetric. 
The effects, possibly detrimental, of the off-diagonal terms could be managed with additional cyber-physical forces. 
The intrinsic anisotropy in the effective mass, $A_{FW} \approx 2C_L({L/D})/S_l$, 
is large since $L/D$ is typically large. 
This was the condition we found to be favorable for fast tracking of $G > 1$ rotorcraft. 

While Eq.~\ref{fixed-wing dynamics} is useful to understand the possible fast-tracking of fixed-wing aircraft subjected to small wind gusts, it is a linear model and therefore cannot simultaneously model the Katzmyar effect, which is nonlinear. 
Optimized FT with a virtual mass smaller than real mass may result in similar or greater performance for fixed-wing aircraft than for rotorcraft since both the Katzmyar effect 
and fast-tracking are enhanced by increasing gust responsiveness in some cases. 

For neutrally-buoyant submarines, blimps, or ships, an equation of a similar form to Eq.~\ref{fixed-wing dynamics} applies, 
but the off-diagonal terms are smaller or zero since these vehicles do not generate lift. 
Only birds and fixed-wing aircraft have the ability to simultaneously take advantage of the Katzmayr and fast-tracking mechanisms, so it is plausible that they will outperform other vehicle types.

\section{Conclusions}
We analyze a model of rotorcraft flight that is identical to the one for particles settling through turbulence. 
Particles with a Stokes number and a settling parameter of order one settle more quickly through turbulence than through quiescent fluid, which is called fast tracking. 
With cyber-physical forcing proportional to acceleration as well as a flight speed adjustment, rotorcraft can mimic any particle settling behavior including fast tracking, and so can extract energy from turbulence to fly with less energy or more quickly. 
By simulating mass anisotropy, the forcing can match not only any particle settling behavior, 
but also behaviors even more favorable than those produced by particles. 
Incidentally, we found that the limiting behavior for large mass anisotropy strengthens previous conclusions made about the importance of the sweeping mechanism to particle settling. 

We show that energy consumed by a rotorcraft in turbulence can be estimated, 
given a certain thrust and inertia, from empirical observations of a vehicle's mean velocity and acceleration variance, and without need for any other information. 
We use this relationship to optimize the parameters of a cyber-physical forcing. 
The optimized forcing reduces energy consumption and flight time in ways that we quantify in a turbulence model. 

We find that energy can be harvested from turbulence by amplifying disturbances to a straight trajectory, and so by increasing flight path length. 
In contrast to existing methods, the principle works for any vehicle traversing turbulence, including fixed-wing aircraft and volant lifeforms, 
and works without knowledge of the flow field. 
In a turbulence model, the advantages in energy consumption and flight time are up to 10\% and 40\%, respectively. 
The results suggest increased performance in real turbulence beyond those we calculated in a turbulence model, and especially for faster and heavier vehicles. 

\subsection*{Acknowledgements}
We are grateful to Profs. S. Ferrari, E. Fisher and Z. Warhaft. 
\subsection*{Declaration of interests}
 The authors report no conflict of interest.

\bibliographystyle{jfm}

\end{document}